\documentclass[[journal, 11pt, onecolumn]{IEEEtran}

\usepackage{subcaption}
\usepackage{amsmath}
\usepackage{amsthm}
\usepackage{dsfont}
\usepackage{comment}
\usepackage{graphicx}
\usepackage{amssymb}
\usepackage{epsfig}
\usepackage{authblk}
\usepackage[bbgreekl]{mathbbol}
\usepackage{floatrow}
\usepackage[font=footnotesize, labelfont=bf]{caption}

\usepackage[all]{xy}
\usepackage[bbgreekl]{mathbbol}
\usepackage{pstricks,pst-node,pst-text,pst-3d}
\usepackage{pst-node}
\newpsstyle{Cdet}{}
\newpsstyle{Cmisdet}{fillcolor=gray!40,fillstyle=solid}

\usepackage[square, sort, numbers]{natbib}
\usepackage[ruled,vlined]{algorithm2e}

\begin{document}
\title{Bayesian tracking and parameter learning for non-linear multiple target tracking models}
\author{Lan Jiang$^{* a}$,
        Sumeetpal S. Singh$^a$, 
        Sinan Y{\i}ld{\i}r{\i}m$^{b}$\\
        \thanks{$a$: lj281@cam.ac.uk (L. Jiang),  sss@cam.ac.uk (S.Singh),  Department of Engineering, University of Cambridge, UK}
        \thanks{$b$:  sy13390@bristol.ac.uk (S. Y{\i}ld{\i}r{\i}m), School of Mathematics, University of Bristol, UK}
        \thanks{L. Jiang's and S.S. Singh's research is funded by the Engineering and Physical Sciences Research Council (EP/G037590/1) whose support is gratefully acknowledged.}
        }


\maketitle


\begin{abstract}
We propose a new Bayesian tracking and parameter learning algorithm for non-linear non-Gaussian multiple target tracking (MTT) models. We design a Markov chain Monte Carlo (MCMC) algorithm to sample from the posterior distribution of the target states, birth and death times, and association of observations to targets, which constitutes the solution to the tracking problem, as well as the model parameters. In the numerical section, we present performance comparisons with several competing techniques and demonstrate significant performance improvements in all cases.

\end{abstract}

\section{Introduction}
The multiple target tracking (MTT) problem is to infer, as accurately as possible, the states or tracks of multiple moving objects from noisy measurements. The problem is made difficult by the fact that the number of targets is unknown and changes over time due to the birth of new targets and the death of existing ones. Moreover, objects are occasionally undetected, false non-target generated measures (clutter) may be recorded and the association between the targets and the measurements is unknown.

Given observations recorded over a length of time, say from time $1$ to $n$, our aim is to jointly infer the target tracks and the MTT model parameters. We adopt a Bayesian approach and our main contribution in this paper is a new Markov chain Monte Carlo (MCMC) algorithm to sample from the MTT posterior distribution, which is a trans-dimensional distribution with mixed continuous and discrete variables. The discrete variables are comprised of the number of targets, birth and death times, and association of observations to targets, while the continuous variables are individual target states and model parameters.

For a linear Gaussian MTT model (see Section \ref{sec: Comparison with MCMC-DA for the  linear Gaussian  MTT model}) an MCMC method for tracking, excluding parameter learning, was proposed in \cite{Oh_et_al_2009}. This MCMC algorithm samples in a much smaller space than we have to since  the continuous valued target states can be integrated out analytically; i.e. it amounts to sampling a probability mass function on a discrete space. (Their method is referred to as MCMC-DA hereinafter.)  However, this model reduction cannot be done for a general non-linear and non-Gaussian MTT model, so the sampling space has to be enlarged to include the continuous state values of the targets. Despite this, our new algorithm is efficient in that it approaches the performance of MCMC-DA for the linear Gaussian MTT model, which will be  demonstrated  in the numerical section.

An MCMC algorithm for tracking in a non-linear non-Gaussian MTT model, but excluding parameter learning, was also recently proposed by \cite{vu2014particle}. Their method follows the MCMC-DA technique of \cite{Oh_et_al_2009}  closely. Although the likelihood of the non-linear non-Gaussian MTT model is not available when the continuous valued states of the targets are integrated out, an unbiased estimate of it can be obtained using a particle filter. The Metropolis-Hastings algorithm can indeed be applied as long as the likelihood of the Bayesian posterior can be estimated in an unbiased fashion and this has been the subject of many recent papers in Bayesian computation; e.g. see \cite{ andrieu2010particle, jasra2013alive, chopin2013smc2}.  
This property is exploited in  \cite{vu2014particle}  and their MCMC sampler for tracking is essentially the MCMC-DA method combined with an unbiased estimate of the likelihood of the reduced model (i.e. continuous states integrated out) which is given by the particle filter. (In the literature on Bayesian computation, this algorithm is known as the Particle Marginal Metropolis Hastings (PMMH); see \cite{andrieu2010particle} for an extensive discussion in a non-MTT context.) Although appealing because it is simple to implement,  the method in  \cite{vu2014particle} can result in an inefficient sampler as we show when comparing with our method. This is because the likelihood estimate has a high variance and this will reduce the overall average acceptance probability of the algorithm. When static parameters are taken into account, which  \cite{vu2014particle} did not do, the variance problem becomes far worse as many products that form the MTT likelihood would have to be simultaneously unbiasedly estimated for the acceptance probability of every proposed parameter change. An elegant solution to this problem is the Particle Gibbs (PGibbs) algorithm of \cite{andrieu2010particle}  for parameter learning in state-space models; we extend this technique to the MTT model.

Our MCMC algorithm for tracking and parameter learning  is a batch method and is suitable for applications where real-time tracking is not essential; e.g. the recent surge in the use of tracking in Single Molecule Fluorescence Microscopy \cite{serge2008dynamic, Yoon_and_Singh_2008}. However, our technique can be incorporated into existing  online trackers (e.g., the  Multiple Hypotheses Tracking (MHT) algorithm \cite{cox1996efficient}, the Joint Probabilistic Data Association Filter (JPDAF) \cite{Bar-Shalom_and_Fortmann_1988}, and the Probability Hypothesis Density (PHD) filter \cite{Mahler_2003, vo2006gaussian}) to correct  past tracking errors in light of new observations as well as  for learning the parameters.  There are numerous ways to effect this,  for example, by applying MCMC to tracks within a fixed window of time, which is a technique frequently used in the \textit{particle filtering} literature for online inference in state-space models. See \cite{del2006sequential, chopin2002sequential} for more discussions on this. Note that, on-line trackers mentioned above normally  ignore parameter learning problem with a few exceptions discussed in \cite{yildirim2014calibrating} where an online maximum likelihood method was proposed for calibrating linear Gaussian MTT model.

Additional contributions of this paper are several interesting comparisons with existing methods.
(i) To quantify the loss of efficiency of our new algorithm compared to MCMC-DA  \cite{Oh_et_al_2009}  that works on a reduced sampling space, we compare them directly for linear Gaussian MTT model,  
and show that we do indeed perform almost comparably to MCMC-DA.
(ii) A comparison with  \cite{vu2014particle}  is given to show that our technique outperforms theirs with much less particles. 
(iii) To demonstrate improvements over online tracking, we present a comparison with the MHT algorithm \cite{cox1996efficient}. As mentioned before, our technique is not a competitor to online tracking but can be incorporated into such trackers to correct past errors.
(iv) We compare our parameter estimates with those obtained by the approximate maximum likelihood technique in \cite{Singh_et_al_2011}
which is built on  the Poisson approximation of the  likelihood. While ours is Bayesian, there should be,  at least,  agreement between the maximum likelihood estimate and the mode of the posterior. We show that 
some parameter estimates obtained by \cite{Singh_et_al_2011} are significantly biased.

The remainder of the paper is organised as follows. In Section \ref{sec: Multiple target tracking model}, we describe the MTT model and formulate the Bayesian target tracking and static parameter estimation problems for the MTT model. In Section \ref{sec: Tracking with known parameters}, we propose a new MCMC tracking algorithm that combines a novel extension of MCMC-DA algorithm to non-linear MTT models with a particle Gibbs move for effectively refreshing the samples for target tracks. In Section \ref{sec: Static parameter estimation}, we show how to do Bayesian static parameter estimation based on the MCMC tracking algorithm presented in Section \ref{sec: Tracking with known parameters}. Numerical examples are shown in Section \ref{sec: Numerical examples} for the comparisons mentioned above. 

\section{Multiple target tracking model} \label{sec: Multiple target tracking model}
The hidden Markov model (HMM), or the state-space model (SSM), is a class of models commonly used for modelling the physical dynamics of a \emph{single} target. In an HMM, a latent discrete-time Markov process $\{ X_{t} \}_{t \geq 1}$ is observed through a process $\{Y_{t} \}_{t \geq 1}$ of observations such that
\begin{align} \label{eq: state-space equations}
\begin{split}
&X_{1} \sim \mu_{\psi}(\cdot), \quad X_{t} | (X_{1:t-1} = x_{1:t-1}) \sim f_{\psi} ( \cdot | x_{t-1}) \\
&Y_{t} |  \left( \left\{ X_{i} =  x_{i} \right\}_{i \geq 1}, \left\{ Y_{i} = y_{i} \right\}_{i \neq t} \right) \sim g_{\psi} ( \cdot | x_{t}).
\end{split}
\end{align}
where $X_{t} \in \mathcal{X} \subset \mathbb{R}^{d_{x}}$, $Y_{t} \in \mathcal{Y} \subset \mathbb{R}^{d_{y}}$, $d_{x} > 0$ and $d_{y} > 0$ are the dimensions of the state and observation. In this paper, 
a random variable (r.v.) is denoted by a capital letter, while its realisation is denoted by a small case. We call  $\mu_{\psi}$, $f_{\psi}$, $g_{\psi}$  the initial, transition, and measurement densities respectively (resp.), and they are parametrised by a real valued vector $\psi \in \Psi \subset \mathbb{R}^{d_{\psi}}$.

In an MTT model, the state and the observation at each time are the random finite sets (we use bold letters to denote sets):
\[
\mathbf{X}_{t} = \left( X_{t, 1}, X_{t, 2}, \ldots, X_{t, K^{x}_{t}} \right), \mathbf{Y}_{t} = \left( Y_{t, 1}, Y_{t, 2}, \ldots, Y_{t, K^{y}_{t}} \right).
\]
Each element of $\mathbf{X}_{t}$ is the state of an individual target. The number of targets $K^{x}_{t}$ under surveillance changes over time due to the death of existing targets and the birth of new targets.  Independently from other targets, a target survives to the next time with survival probability $p_{s}$ and its state evolves according to the transition density $f_{\psi}$, otherwise it `dies'. In addition to the surviving targets, new targets are `born' from a Poisson process with density $\lambda_{b}$ and each of their states is initialised by sampling from the initial density $\mu_{\psi}$. The hidden states of the new born targets and surviving targets from time $t-1$ make up $\mathbf{X}_{t}$. We assume that at time $t=1$ there are only new born targets, i.e.\ no surviving targets from the past. Independently from other targets, each target in $\mathbf{X}_{t}$ is detected and generates an observation according to observation density $g_{\psi}$ with probability $p_{d}$. In addition to observations generated from detected targets, false measurements (clutter) can appear from a Poisson process with the density $\lambda_{f}$ and are uniformly distributed over $\mathcal{Y}$. We denote by $\mathbf{Y}_{t}$ the superposition of clutter and measurements of the detected targets.

\subsection{The law of MTT model} \label{sec: The law of MTT model}
In the following, we give a description of  the generative model of the MTT problem, where $\mathbf{X}_t, \mathbf{Y}_t$ are treated as ordered sets for convenience. A series of r.v.'s are now defined to give a precise characterisation of the MTT model.
Let $C^{s}_{t}$ be a $K^{x}_{t-1} \times 1$ vector of $1$'s and $0$'s where $1$'s indicate survivals and $0$'s indicate deaths of targets from time $t-1$. For $ i = 1:K^{x}_{t-1}$,
\begin{equation}
C^{s}_{t}(i) = \begin{cases}
      1 & \text{$i$'th target at time $t-1$ survives to time $t$} \\
      0 & \text{$i$'th target at time $t-1$ does not survive to $t$}
\end{cases}. \nonumber
\end{equation}
Denote $K_t^s$ the number of surviving targets at time $t$, and $K_t^b$ the number of `birth' at time $t$. We have
\[K^{s}_{t} = \sum_{i = 1}^{K^{x}_{t-1}} C^{s}_{t}(i),\quad K_t^x=K_t^s+K_t^b.\]
At time $t$, the surviving targets from time $t-1$ are re-labeled as $X_{t, 1}, \ldots, X_{t, K^{s}_{t}}$, and the newly born targets are denoted as $X_{t, K^{s}_{t}+1}, \ldots, X_{t, K^{x}_{t}}$ (according to certain numbering rule specified by users as will be addressed shortly). The order of the surviving targets at time $t$ is determined  by their ancestor order at time $t-1$. Specifically, we define the $K^{s}_{t} \times 1$ ancestor vector $I^{s}_{t}$ for $X_{t,i},\; i=1:K_t^s$,
\begin{equation}
I^{s}_{t}(i) = \min \big\{ k: \sum_{j = 1}^{k} C^{s}_{t}(j) = i \big\}, \quad i = 1: K^{s}_{t}. \nonumber
\end{equation}
Note that $I_{t}^{s}(i)$ denotes the ancestor of target $i$ from time $t-1$, i.e., $X_{t-1, I^{s}_{t}(i)}$ evolves to $X_{t, i}$ for $i = 1: K^{s}_{t}$. Next, we define $I^{d}_{t}$ to be a $K^{x}_{t} \times 1$ vector showing the target to measurement association at time $t$. For  $j = 1:K^{x}_{t}$,
\begin{equation}
I^{d}_{t}(j) = \begin{cases}
      k & \text{if $X_{t, j}$ generates $Y_{t, k}$}, \\
      0 & \text{$X_{t, j}$ is not detected}.
\end{cases} \nonumber
\end{equation}
Denote $K_t^d$ the number of detected targets at time $t$, and $K_t^f$ the number of false measurements at time $t$. We have
\[K_{t}^{d} = \# \{ j : I_{t}^{d}(j) > 0 \}, \quad K_{t}^{y} = K_{t}^{f}+K_{t}^{d}.\]
where $\#$ denotes the cardinality of the set. Sampling from the prior of $I_t^d$, amounts to first sampling a binary  $K_t^x \times 1$ detection vector whose element  is an independent and identically distributed (i.i.d.) Bernoulli r.v. with success parameter $p_d$ (to decide which targets are detected, i.e, indices of non-zero entries in $I_t^d$), then  sample a $K_t^d\times 1$ association vector to determine the association between detected targets and observations uniformly from all $k_t^d$-permutations of $k_t^y$, i.e, with probability $\frac{k_t^f!}{k_t^y!}$  (to decide specific values for non-zeros entires of $I_t^d$).

The main difficulty in the MTT problem is that we do not know birth-death times of targets, whether they are detected or not, and which measurement point in $\mathbf{Y}_{t}$ is associated to which detected target in $\mathbf{X}_{t}$. Now we define \textit{data association} 
\begin{equation}
Z_{t} = \big(C^{s}_{t}, K_t^b, K^{f}_{t},I^{d}_{t}\big)\label{eq:DA}
\end{equation}
to be the collection of the above mentioned unknown r.v.'s at time $t$, and
\begin{equation}
\theta = (\psi, p_{s}, p_{d}, \lambda_{b}, \lambda_{f}) \in \Theta = \Psi \times [0, 1]^{2} \times [0, \infty)^{2}\label{eq:theta}
\end{equation}
be the vector of the MTT model parameters. Assuming  survival and detection probabilities are state independent,   we can write down the MTT model described literally above as
\begin{align}
&\hspace{-0.15cm}p_{\theta}(z_{1:n}) =\prod_{t = 1}^{n} \biggl(p_{s}^{k^{s}_{t}} (1 - p_{s})^{k^{x}_{t-1} - k^{s}_{t}}  \mathcal{PO}(k^{b}_{t} ; \lambda_{b}) 
 \mathcal{PO}(k^{f}_{t}; \lambda_{f})\, p_{d}^{k^{d}_{t}} (1 - p_{d})^{ k^{x}_{t} - k^{d}_{t} }\,\frac{k_t^f!}{k_t^y!}\biggr) \label{eq: density of z}\\
&p_{\theta}(\mathbf{x}_{1:n} | z_{1:n} ) = \prod_{t = 1}^{n}\, \biggl[ \prod_{ j=1 }^{k_t^s} f_{\psi}(x_{t, j} | x_{t-1, i^{s}_{t}(j)})
 k_t^b! \mathbf{1}_{A}(x_{t,k_t^s+1:k_t^x})\prod_{j=k_t^s+1}^{k_t^x}\mu_{\psi}(x_{t, j}) \biggr]\label{eq: density of x given z}\\
&p_{\theta}(\mathbf{y}_{1:n} | \mathbf{x}_{1:n}, z_{1:n} ) = \prod_{t = 1}^{n}  \biggl[ \left\vert \mathcal{Y} \right\vert^{-k^{f}_{t}} \hspace{-0.3cm}\prod_{j: i_{t}^{d}(j) > 0}\hspace{-0.2cm} g_{\psi}(y_{t, i_{t}^d(j)} | x_{t, j})  \biggr]. \label{eq: density of y given x and z}
\end{align}
Here $a_{i:j}, \, i\leq j$ is used to denote a finite sequence $\{a_i,a_{i+1}\ldots a_j\}$,  $\mathcal{PO}(k; \lambda)$ denotes the probability mass function of the Poisson distribution with mean $\lambda$, 
$\left\vert \mathcal{Y} \right\vert$ is the volume (the Lebesgue measure) of $\mathcal{Y}$, and $\mathbf{1}_A$ is the indicator function of the numbering rule  $A$ for the new born targets (e.g, if new-borns are ordered in an ascending order of the first component, then $A$ is  the set of  states satisfying $x_{t,k_t^s+1}(1)<\cdots<x_{t,k_t^x}(1)$).\footnote{$A$ is introduced here to avoid the labelling ambiguity of new born targets. The labelling ambiguity also arises in other areas, e.g. Bayesian inference of mixture distributions; see \cite{jasra2005markov} for more details.} So the joint density of all the variables of the MTT is
\[
p_{\theta}(z_{1:n}, \mathbf{x}_{1:n}, \mathbf{y}_{1:n}) =  p_{\theta}(z_{1:n}) p_{\theta}(\mathbf{x}_{1:n} | z_{1:n}) p_{\theta}(\mathbf{y}_{1:n} | \mathbf{x}_{1:n}, z_{1:n}).
\]
Finally, the marginal likelihood of the data $\mathbf{y}_{1:n}$ is given by
\[
p_{\theta}(\mathbf{y}_{1:n})\! =\! \sum_{z_{1:n}} p_{\theta}(z_{1:n})\hspace{-0.1cm} \int\hspace{-0.1cm} p_{\theta}(\mathbf{y}_{1:n} | \mathbf{x}_{1:n}, z_{1:n}) p_{\theta}(\mathbf{x}_{1:n} | z_{1:n}) d \mathbf{x}_{1:n}.
\]

\subsection{Two equivalent mathematical descriptions for MTT} \label{sec: Two equivalent notations for MTT}
Note that, conditional on $Z_{1:n}$, $( \mathbf{X}_{1:n}, \mathbf{Y}_{1:n} )$ may be regarded as a collection of  HMMs (with different starting and ending times and possible missing observations) and observations which are not relevant to any of these models. In the MTT terminology, each HMM corresponds to a target, starting and ending times of HMMs correspond to birth and death times of those targets, and missing and irrelevant observations correspond to mis-detections and clutter.

Note that, each target has a distinct label  $k\in\{1,\ldots, K\}$ where $K = \sum_{t = 1}^{n} k_{t}^{b}$, which is determined by its birth time and the numbering of its initial state at the birth time (dependent on the numbering rule). Let $t_{b}^{k}$ and $t_{d}^{k}$ be the birth and death time of the target with label $k$, and denote its trajectory as
\[
\hat{\mathbf{X}}^{(k)}=(\hat{X}_{1}^{(k)},\ldots, \hat{X}_{l_k}^{(k)}),\; \hat{\mathbf{Y}}^{(k)}= (\hat{Y}_{1}^{(k)},\ldots, \hat{Y}_{l_{k}}^{(k)})
\]
where  $\hat{X}_{i}^{(k)}$ is the $i$-th state of target $k$;  $\hat{Y}_{i}^{(k)}$ is the observation generated by $\hat{X}_{i}^{(k)}$ provided detection, otherwise we take $\hat{Y}_{i}^{(k)} = \varnothing$;  $l_{k} = t_{d}^{k} - t_{b}^{k}$ is its life span. In particular, $\hat{\mathbf{X}}^{(k)}, \hat{\mathbf{Y}}^{(k)}$ form a HMM with initial and state transition densities $\mu_{\psi}$ and $f_{\psi}$ and observation density $g_{\psi}$ as in \eqref{eq: state-space equations} with the convention that $g_{\psi}(\varnothing | x)  = 1,\; x \in \mathcal{X}$ to handle mis-detections. In addition, we define $\hat{\mathbf{Y}}^{(0)}$ that contains all irrelevant observations during time $1:n$ with $\hat{\mathbf{X}}^{(0)}=\varnothing$. 

To recover $(Z_{1:n}, \mathbf{X}_{1:n}, \mathbf{Y}_{1:n})$ from $ \{ \hat{\mathbf{X}}^{(k)}, \hat{\mathbf{Y}}^{(k)}\}_{k=0}^K $, we also need to know $\hat{Z}^{(k)}$ which contains\footnote{We can write $\hat{Z}^{k}=(t_b^k, t_d^k, I_y^k)$ where $I_y^{k}$ is a $l_k\times 1$ vector with $I_y^k (i)$ being the index of  $\hat{Y}_i^{(k)}$ in  $\mathbf{Y}_{t}$ (the collection of all observations at its appearing time $t$) if $\hat{Y}_i^{(k)}\neq \varnothing$,  otherwise $I_y^k (i)= 0$.} the information of the birth time, the death time and the indices of measurements assigned to target $k$ for $k=1:K$.  $\hat{Z}^{(0)}$ is defined for clutter so that it contains all clutter's appearance times and their corresponding measurement indices.
The point we want to make here is that given ordering rule $A$ for new born targets, we have a one-to-one mapping between the two equivalent  descriptions of the MTT model, i.e.\
\begin{equation}
Z_{1:n}, \mathbf{X}_{1:n}, \mathbf{Y}_{1:n} \Leftrightarrow \{ \hat{Z}^{(k)},\hat{\mathbf{X}}^{(k)}, \hat{\mathbf{Y}}^{(k)}\}_{k=0}^K. \label{eq:TwoNotations}
\end{equation}
In Figure \ref{fig: MTT}, we give a realisation of the MTT model to illustrate the r.v.'s introduced in both descriptions  and show the correspondence between these two descriptions. It can be seen that  each target (HMM) evolves and generates observations independently, with the only dependancy  introduced by the target labels dependent on the numbering rule. 
\begin{figure}[h]
\begin{center}
\small
\psset{arrows=->,mnode=circle,linewidth=0.0pt, colsep=0.08cm,rowsep=0.01cm}
\begin{psmatrix}
[name=X11,  style=Cdet] $X_{1, 1}$ & & [name=X21,  style=Cmisdet] $X_{2, 1}$ & & [name=X31,  style=Cdet] $X_{3, 1}$  & & [name=X41,  style=Cmisdet] $X_{4, 1}$  \\
& [name = Y11, mnode=r] \psframebox{$Y_{1, 3}$} & & & & [name=Y31, mnode=r] \psframebox{$Y_{3, 2}$} \\
[name=X12,  style=Cdet]$X_{1, 2}$ & & [name=X22,  style=Cdet] $X_{2, 2}$ & & [name=X32,  style=Cdet]$X_{3, 2}$  & &  [name=X42,  style=Cdet]$X_{4, 2}$ \\
& [name=Y12, mnode=r] \psframebox{$Y_{1, 1}$} & & [name=Y22, mnode=r] \psframebox{$Y_{2, 1}$} & & [name=Y32, mnode=r] \psframebox{$Y_{3, 3}$}  & & [name=Y42, mnode=r] \psframebox{$Y_{4, 1}$} \\
[name=X13,  style=Cmisdet]$X_{1, 3}$ & & [name=X23,  style=Cdet]$X_{2, 3}$ & &[name=X43,  style=Cmisdet]$X_{3, 3}$ \\
& [name=Y13, mnode=r] \psframebox[linestyle=dashed, linewidth=1pt]{$Y_{1, 2}$} & & [name=Y23, mnode=r] \psframebox{$Y_{2, 2}$} & & [name=Y33, mnode=r] \psframebox[linestyle=dashed, linewidth=1pt]{$Y_{3, 1}$}
\ncline[linewidth = 1pt]{X11}{X21} \ncline[linewidth = 1pt]{X12}{X22} \ncline[linewidth = 1pt]{X13}{X23}
\ncline[linewidth = 1pt]{X21}{X31}\ncline[linewidth = 1pt]{X22}{X32}
\nccurve[angleA=0,angleB=-100, linewidth = 1pt]{X32}{X41} \nccurve[angleA=0,angleB=-100, linewidth = 1pt]{X33}{X42}
\ncline[linewidth = 0.5pt]{X11}{Y11} \ncline[linewidth = 0.5pt]{X12}{Y12}
\ncline[linewidth = 0.5pt]{X22}{Y22} \ncline[linewidth = 0.5pt]{X23}{Y23}
\ncline[linewidth = 0.5pt]{X31}{Y31} \ncline[linewidth = 0.5pt]{X32}{Y32}
\ncline[linewidth = 0.5pt]{X42}{Y42}
\end{psmatrix}
\caption{A realisation from the  MTT model: states of a targets are connected with arrows and with their observations when detected. Undetected targets are coloured grey, and false measurements are in dashed lines.  For this example, 
\newline $c^s_{1:4}=\left([ ], [1,1,1], [1,1,0], [ 0, 1, 0]\right)$,
 $i^{s}_{1:4} = \left( \left[  0,0,0\right],  \left[ 1, 2, 3 \right],  \left[ 1, 2\right],  \left[ 2\right]\right)$,  $k_{1:4}^b=(3,0,1,1)$, 
 \newline $k^{f}_{1:4} = \left( 1, 0, 1, 0 \right),i^{d}_{1:4} = \left( \left[ 3, 1,0 \right],  \left[0, 1, 2 \right],  \left[ 2, 3,0 \right],  \left[0,1 \right] \right)$;
\newline$\hat{\mathbf{x}}^{(1)}=(x_{1,1},x_{2,1},x_{3,1})$, $\hat{\mathbf{y}}^{(1)}=(y_{1,3},\varnothing, y_{3,2})$, $\hat{z}^{(1)}=(1,4,[3,0,2])$;
\newline $\hat{\mathbf{x}}^{(2)}=(x_{1,2}, x_{2,2}, x_{3,2},x_{4,1})$, $\hat{\mathbf{y}}^{(2)}=(y_{1,1},y_{2,1}, y_{3,3},\varnothing)$, $\hat{z}^{(2)}=(1,5,[1,1,3,0])$, 
\newline $\hat{\mathbf{x}}^{(3)}=(x_{1,3}, x_{2,3})$, $\hat{\mathbf{y}}^{(3)}=(\varnothing, y_{2,2})$, $\hat{z}^{(3)}=(1,3,[0,2])$; 
\newline $\hat{\mathbf{x}}^{(4)}=(x_{3,3})$, $\hat{\mathbf{y}}^{(4)}=(\varnothing)$, $\hat{z}^{(4)}=(3,4,[0])$; \;
$\hat{\mathbf{x}}^{(5)}=(x_{4,2})$, $\hat{\mathbf{y}}^{(5)}=(y_{4,1})$, $\hat{z}^{(5)}=(4,5,[1])$.
}\label{fig: MTT}
\end{center}
\end{figure}

Although it is more straightforward  to write down the MTT probability model in terms of the first description, see \eqref{eq: density of z}-\eqref{eq: density of y given x and z}, the second description here is indispensable for our  MCMC moves where we first propose change to $\hat{Z}^{(k)},\hat{\mathbf{X}}^{(k)}$ for some target $k$ or a set of targets, then we get the unique $Z_{1:n}, \mathbf{X}_{1:n}$ based on the equivalence of these two descriptions.

\subsection{Bayesian tracking and parameter estimation for MTT} \label{sec: Bayesian smoothing and parameter estimation for MTT}
There are two main problems we are interested in this paper: assuming $\theta$ is known, the first one is to estimate the data association and the states of the targets given the observations $\mathbf{y}_{1:n}$. This problem is formalised as estimating the posterior distribution
\begin{align}
p_{\theta}(z_{1:n}, \mathbf{x}_{1:n} | \mathbf{y}_{1:n}) = \frac{p_{\theta}(z_{1:n}, \mathbf{x}_{1:n}, \mathbf{y}_{1:n})}{p_{\theta}(\mathbf{y}_{1:n})} \label{eq: MTT posterior distribution}
\end{align}
where $p_{\theta}(\mathbf{y}_{1:n})$ serves as a normalising constant not depending on $(z_{1:n}, \mathbf{x}_{1:n})$. We present a novel MCMC method which samples from the posterior distribution \eqref{eq: MTT posterior distribution} for non-linear  MTT models in Section \ref{sec: Tracking with known parameters}.

The second problem we are interested is the static parameter estimation problem, that is estimating $\theta$ from the data $\mathbf{y}_{1:n}$. We regard $\theta$ as a r.v. taking values in $\Theta$ with a prior density $\eta(\theta)$, and our goal is to estimate the posterior distribution of $\theta$ given data, that is
\begin{equation}
p(\theta | \mathbf{y}_{1:n}) \propto \eta(\theta) p_{\theta}(\mathbf{y}_{1:n}) \label{eq: posterior of theta given y}
\end{equation}
which is intractable for MTT models in general. In Section \ref{sec: Static parameter estimation}, we extend our MCMC tracking method in Section \ref{sec: Tracking with known parameters} to get samples $(\theta, \mathbf{x}_{1:n}, z_{1:n})$ from the joint posterior distribution $p(\theta, z_{1:n}, \mathbf{x}_{1:n} |\mathbf{y}_{1:n})$. 

\section{Tracking with known parameters} \label{sec: Tracking with known parameters}
In this section we assume the parameter $\theta$ of the MTT model is known and we want to estimate the posterior density $p_{\theta}(\mathbf{x}_{1:n}, z_{1:n}|\mathbf{y}_{1:n})$ defined in \eqref{eq: MTT posterior distribution}.

For a linear Gaussian MTT model, one can consider the following factorisation of the posterior density
\[
p_{\theta}(z_{1:n}, \mathbf{x}_{1:n} | \mathbf{y}_{1:n}) = p_{\theta}(z_{1:n} | \mathbf{y}_{1:n}) p_{\theta}(\mathbf{x}_{1:n} | z_{1:n}, \mathbf{y}_{1:n})
\]
and concentrate on sampling from $p_{\theta}(z_{1:n} | \mathbf{y}_{1:n}) \propto p_{\theta}(z_{1:n}) p_{\theta}(\mathbf{y}_{1:n} | z_{1:n})$, as  $p_{\theta}(\mathbf{y}_{1:n} | z_{1:n})$, the likelihood of the data given the data association, can be calculated exactly. Similarly, once we have samples for $z_{1:n}$, $p_{\theta}(\mathbf{x}_{1:n} | z_{1:n}, \mathbf{y}_{1:n})$ can be calculated exactly for every sample of $z_{1:n}$.\footnote{Strictly speaking, the closed forms are available when we ignore the ordering  rule here.} This is indeed the case for the MCMC-DA algorithm of \cite{Oh_et_al_2009}, which is essentially an MCMC algorithm for sampling from $p_{\theta}(z_{1:n} | \mathbf{y}_{1:n})$. However, when the MTT model is non-linear, which is the case in this paper, MCMC-DA is not applicable since $p_{\theta}(\mathbf{y}_{1:n} | z_{1:n})$ is not available.
\cite{vu2014particle} proposed to circumvent this by using an unbiased estimator $\hat{p}_{\theta}(\mathbf{y}_{1:n} | z_{1:n})$ in place of $p_{\theta}(\mathbf{y}_{1:n} | z_{1:n})$, which is obtained by running a particle filter for each target.  
This is essentially the PMMH algorithm of \cite{andrieu2010particle} applied to the MTT problem.
However, this strategy mixes slowly due to the variance of the estimate of $p(\mathbf{y}_{1:n}|z_{1:n})$, especially when the number of particles is small, which is demonstrated in Section \ref{sec: Comparison with MCMC-DA for the  linear Gaussian  MTT model}. It is also not efficient since $\mathbf{X}_{1:n}$ is  only a by-product of the PMMH algorithm, and not used to propose the change of data association $Z_{1:n}$.
In this paper, we first design  an efficient sampler to change $Z_{1:n}$ and $X_{1:n}$ together based on the old samples to avoid the variance problem encountered in  the PMMH when the particle number is small. Then, we refresh  $\mathbf{X}_{1:n}$ by applying the particle Gibbs (PGibbs) algorithm proposed in \cite{andrieu2010particle} to accelerate mixing.

This section documents our MCMC algorithm for sampling $(z_{1:n}, \mathbf{x}_{1:n})$ jointly from \eqref{eq: MTT posterior distribution}. Before going into the details, it will be useful to have an insight into the distribution in \eqref{eq: MTT posterior distribution}. Notice that the dimension of $\mathbf{X}_{1:n}$ is proportional  to  $\sum_{t = 1}^{n} K^{x}_{t}$ which is determined by the data association $Z_{1:n}$. Therefore, the posterior distribution in \eqref{eq: MTT posterior distribution} is  trans-dimensional and the standard Metropolis-Hastings (MH) algorithm 
is not applicable for this distribution.

A general method for sampling from a trans-dimensional distribution is the reversible jump MCMC (RJ-MCMC) algorithm of \cite{green1995reversible}. Assume we have the target distribution $\pi(m, x_{m})$ where
$m$ is discrete, and $x_m$ is a vector with dimension $d_{m}$ that changes with $m$. Here, $m$ can be considered as a model index, whose dimension $d_m$ is not necessarily different from $d_{m'}$ for  $m'\neq m$.
To move a sample $(m, x_{m})$ from $\pi(m, x_{m})$ to a subspace with a higher dimension,  we can first propose $(m', u_{m,m'})\sim q(\cdot| m, x_m)$, where $m'$ is the model index such that $d_{m'}>d_m$, and $u_{m,m'}\in \mathcal{R}^{d_{m,m'}}$ are extra continuous r.v.'s such that $d_{m'}=d_m+d_{m,m'}$ (dimension matching). Finally  the candidate sample is given by a \textit{bijection}: $x_{m'}=\beta_{m,m'}(x_m, u_{m,m'})$. 
For the reverse move, with probability $q(m|m', x_{m'})$ propose to move to subspace $m$, and use the bijection $\beta_{m',m}=\beta_{m,m'}^{-1}$ to get $(x_{m}, u_{m,m'})=\beta_{m,m'}^{-1}(x_{m'})$.
The acceptance probability 
for the proposed sample $(m', x_{m'})$ is $\alpha(m',x_{m'}; m, x_m)=\min\{1, r(m',x_{m'}; m, x_m)\}$  where
\begin{equation}
r(m',x_{m'}; m, x_m)=\frac{\pi(m', x_{m'})}{\pi(m, x_{m})}\times
 \frac{q(m | m', x_{m'})}{ q(m' , u_{m,m'}| m, x_m) }
 \!\!\left\vert \frac{\partial x_{m'}}{\partial(x_{m}, u_{m, m'})} \right\vert\label{eq:TransMC}
\end{equation}
where the rightmost term is the Jacobian of $\beta_{m,m'}$. The acceptance ratio of the reverse move is 
\begin{equation}
r(m,x_{m}; m', x_{m'})=r(m',x_{m'}; m, x_m)^{-1}.\label{eq:TransMC_reverse}
\end{equation}



In the MTT model, each data association $z_{1:n}$  corresponds to a model index $m$, $\mathbf{x}_{1:n}$ corresponds to the continuous variable $x_m$, and $p_{\theta}(z_{1:n}, \mathbf{x}_{1:n}|\mathbf{y}_{1:n})$ corresponds to $\pi(m, x_m)$.
From this perspective, we can devise a RJ-MCMC algorithm for \eqref{eq: MTT posterior distribution} which has two main parts: (i) MCMC moves that are designed to explore the data association $Z_{1:n}$, followed by (ii) an MCMC move that explores the continuous states $\mathbf{X}_{1:n}$. While the later move aims to explore $\mathbf{X}_{1:n}$ only, we also need to adapt $Z_{1:n}$ to respect the adopted ordering rule $A$ of new born targets.
We present a single iteration of the proposed MTT algorithm in Algorithm \ref{alg: MCMC Tracking} referred to as MCMC-MTT.
\begin{algorithm}[h]
\DontPrintSemicolon
\KwIn{Current sample $(z_{1:n}, \mathbf{x}_{1:n})$, data $\mathbf{y}_{1:n}$, parameter $\theta$, number of inner loops $n_{1}$, $n_{2}$}
\KwOut {Updated sample $(z_{1:n}, \mathbf{x}_{1:n})$}
\For{$j = 1:n_{1}$}
{Update $z_{1:n}, \mathbf{x}_{1:n}$ by one of the MCMC moves  in Algorithm \ref{alg: MCMC across data associations} to explore the data association $Z_{1:n}$}
\For{$j = 1:n_{2}$}
{Update $z_{1:n},\mathbf{x}_{1:n}$ by an  MCMC move (Algorithm \ref{alg: MCMC within data association}) to  explore the continuous state space $\mathbf{X}_{1:n}$} 
\caption{MCMC-MTT}\label{alg: MCMC Tracking}
\end{algorithm}

Algorithm \ref{alg: MCMC Tracking} can be viewed as an extension of MCMC-DA \cite{Oh_et_al_2009}  to the non-linear non-Gaussian case by incorporating  $\mathbf{X}_{1:n}$ into the sampling space. Designing the MCMC kernel for the first loop is demanding and we reserve Section \ref{sec: across DA} for the description of this kernel. The second loop uses a  PGibbs kernel to refresh the samples of $\mathbf{X}_{1:n}$ conditioned on the data association, which is an important factor for fast mixing when we enlarge our sampling space. The PGibbs step is standard  since given the data association, the MTT model can be decoupled into a set of  HMMs (as emphasised by the alternative description introduced in Section \ref{sec: Two equivalent notations for MTT}).

We have found that Algorithm \ref{alg: MCMC Tracking} can  work properly with any initialisation for  $z_{1:n}$, even with the all clutter case, i.e.\ $K_{1:n}^{b} = 0$, hence $K_{1:n}^{f} = k_{1:n}^{y}$, and $\mathbf{X}_{1:n} = \varnothing$, which is a convenient choice when no prior information is available. We generally take $n_{1}$ an order of magnitude larger than $n_{2}$ ($n_2=1$ typically) as the second loop takes more time than the first one.


\subsection{MCMC to explore the data association}\label{sec: across DA}
Algorithm \ref{alg: MCMC across data associations}  proposes a new data association with one of the following six moves at random:
\begin{enumerate}
\item \emph{birth move}: to create a new target and its trajectory;
\item \emph{death move}: to randomly delete an existing target;
\item \emph{extension move}: to randomly extend an existing track;
\item \emph{reduction move}: to randomly reduce an existing track;
\item \emph{state move}: to randomly modify the links between state variables at successive times;
\item \emph{measurement move}: to randomly modify the links between state variables and observation variables.
\end{enumerate}
The first four of the moves change the dimension of $\mathbf{X}_{1:n}$, and hence they will be called trans-dimensional moves where RJ-MCMC needs to be applied. Specifically, the dimension matching here is done by introducing new states or deleting existing ones, and the bijections are such that the Jacobian in \eqref{eq:TransMC} is always $1$. Reversibility is ensured by pairing the birth (resp. extension) move with the death (resp. reduction) move. The last two moves, i.e., the state move and the measurement move, leave the dimension of $\mathbf{X}_{1:n}$ unchanged, so called as dimension-invariant moves, and  a normal MH step can be applied. We will see later that these two moves are self-reversible, i.e.,\ they are paired with themselves. In the following subsection, we describe the essence of each move included in Algorithm \ref{alg: MCMC across data associations}. 

\begin{algorithm}[h]
\DontPrintSemicolon
\KwIn{Current sample $(z_{1:n}, \mathbf{x}_{1:n})$, data $\mathbf{y}_{1:n}$, parameter $\theta$, window parameter $\tau$}
\KwOut {Updated sample $(z_{1:n},\mathbf{x}_{1:n})$}
Sample $j \in \{1, \ldots, 6 \}$ randomly. \;
 \Switch{$j$}
  {\lCase {$1$} {propose $(z'_{1:n}, \mathbf{x}'_{1:n})$ by the \emph{birth move}}
   \lCase {$2$} {propose $(z'_{1:n}, \mathbf{x}'_{1:n})$ by the \emph{death move}}
   \lCase {$3$} {propose $(z'_{1:n}, \mathbf{x}'_{1:n})$ by the \emph{extension move}}
   \lCase {$4$} {propose $(z'_{1:n}, \mathbf{x}'_{1:n})$ by the \emph{reduction move}}
   \lCase {$5$} {propose $(z'_{1:n}, \mathbf{x}'_{1:n})$ by the \emph{state move}}
   \lCase {$6$} {propose $(z'_{1:n}, \mathbf{x}'_{1:n})$ by the \emph{measurement move}}
  }
  Calculate the MCMC acceptance probability for move $j$
  \begin{equation*}
  \alpha_{j}=\min\{1,r_j(z'_{1:n}, \mathbf{x}'_{1:n}; z_{1:n}, \mathbf{x}_{1:n})\}\label{eq: AcptEq}
  \end{equation*}
(See \eqref{eq: AcptRatio_BM}, \eqref{eq: AcptRatio_EM}, \eqref{eq: AcptRatio_SM}, \eqref{eq: AcptRatio_MM}, \eqref{eq:TransMC_reverse} for the calcatuion of $r_j$).\;
Change $z_{1:n}=z'_{1:n},\; \mathbf{x}_{1:n}=\mathbf{x}'_{1:n}$ with probability $\alpha_{j}$, otherwise reject the proposal.
\caption{MCMC moves to update data association}\label{alg: MCMC across data associations}
\end{algorithm}

\subsubsection{Trans-dimensional moves}
Two pairs of moves (\emph{birth/death}, \emph{extension/reduction}) are designed to jump between different dimensions for $\mathbf{X}_{1:n}$.

\paragraph{Birth and death moves}
Assume the current sample of our MCMC algorithm for $Z_{1:n}$ implies $K$ existing targets. We propose a new target with randomly chosen birth and death times and randomly assigned observations from the clutter, i.e.\ observations unassigned to any of the existing targets. We give a sketch of the birth move here.

We first propose a random birth time $t_{b}$ and sample death time $t_d\leq (n+1)$ based on $p_s$ (note, $t_d$ can be changed later during this birth process) for the new target, then extend the trajectory of the target forward in time in a recursive way until $t_{d}$. Each extension step proceeds as follows. Assume the latest observation $y_{p}$ we assigned to the new target is observed at time $t_{p}$. (For the first iteration, $t_p=t_b-1$, $y_p$ takes the mean of the initial position.) We define the time block $B = \{ t_{p} +1, \ldots, \min \{ t_{p} + t_{m}, t_{d}-1 \} \}$ where
\[
t_{m} = \min\{ t: (1-p_{d})^{t} < (1-p_{m}) \},
\]
given a user defined probability $p_{m}$ (close to 1). The logic behind this is that within $\{ t_{p}+1, \ldots, t_{p} + t_{m} \}$ the next measurement would appear with \emph{a priori} probability larger than $p_{m}$.   
Among all the unassigned observations in this time block $B$, we form a set of candidate observations whose distance to $y_{p}$ (which depends on both time and space) is less than a certain threshold set by users. Note $p_m$ should be big enough so that block $B$ contains most possible candidates.
(i) With probability $p_{m}$, we decide that the next observation to be assigned to the new target is located in $B$ and choose it randomly from the set of candidate observations with probability  inversely proportional to the distance to $y_p$, provided that the set is non-empty.
If the set of candidate measurements is empty, however, we terminate the target either at $t_{d}$ if $t_{d} \leq t_{p} + t_{m}$, or at some random time in the block (proposed by taking $p_s, p_d, t_p$ into account) otherwise. The termination time is the final proposed death time $t_{d}$ for the target. (ii) If (i) is not performed, i.e.\ with probability $(1 - p_{m})$, we decide that the target is not detected during the whole block $B$. Then we recommence the process above from the end of the block, unless $t_{d} \leq t_{p} + t_{m}+1$.
We refer to this iterative observation assignment procedure as grouping measurement step, 
at the end of which, we obtain $\hat{z}_b$ containing the birth time, the death time and measurement indices of the new born target, and we denote  $q_{b,\theta}(\hat{z}_b|z_{1:n},\mathbf{y}_{1:n})$ the probability induced in this step. 
 The new target's states $\hat{\mathbf{x}}_b$ are proposed by running unscented Kalman filter (UKF)  \cite{wan2000unscented} followed by backwards sampling \cite{doucet2009tutorial}, which is essentially a Gaussian proposal for the target states (see Appendix A for  more on UKF and backwards sampling). Denote $q_{b, \theta}(\hat{\mathbf{x}}_b|\hat{z}_b,\mathbf{y}_{1:n})$ the probability density induced in this step. The sampled hidden states will serve as dimension matching parameters of the RJ-MCMC algorithm. Given the set $\{\hat{z}^{(k)},\hat{x}^{(k)}\}_{k=1}^K\cup \{\hat{z}_b,\hat{\mathbf{x}}_b\}$, new data association $z'_{1:n}$ can be obtained deterministically by the one-to-one mapping \eqref{eq:TwoNotations} mentioned in section \ref{sec: Two equivalent notations for MTT} according to 
the ordering rule. 
Finally, we get new states $\mathbf{x}'_{1:n}=\beta_{z_{1:n}, z'_{1:n}}(\mathbf{x}_{1:n}, \hat{\mathbf{x}}_b )$, where $\beta_{z_{1:n}, z'_{1:n}}$ is to insert $\hat{\mathbf{x}}_b $ into $\mathbf{x}_{1:n}$
at the corresponding positions indicated by $z'_{1:n}$. The resulting Jacobian is $1$.

The death move, which is the reverse move of the birth move, is done by randomly deleting one of the existing tracks. The acceptance ratio of the birth move  is
\begin{equation}
r_{1}(z'_{1:n},\mathbf{x}'_{1:n}; z_{1:n},\mathbf{x}_{1:n}) = \frac{p_{\theta}(z_{1:n}', \mathbf{x}'_{1:n}, \mathbf{y}_{1:n})}{p_{\theta}(z_{1:n}, \mathbf{x}_{1:n},\mathbf{y}_{1:n})}
\times\frac{q_{d,\theta}(z_{1:n}|z'_{1:n})}{ q_{b, \theta}(\hat{z}_{b} | z_{1:n}, \mathbf{y}_{1:n}) q_{b, \theta}(\hat{\mathbf{x}}_b | \hat{z}_b,\mathbf{y}_{1:n}) }
\label{eq: AcptRatio_BM}
\end{equation}
whose reciprocal is the acceptance ratio for the corresponding death move.
Here,  $q_{d,\theta}(z_{1:n}|z'_{1:n})$ is the probability, induced by the death move.
Note that, $q_{b,\theta}(\hat{z}_{b} | z_{1:n}, \mathbf{y}_{1:n})$ depends on  $p_{s}, p_d$  and the distance between the last assigned observation of the target and all clutter in the next few time steps. Thus, in some sense, the move exploits a pseudo-posterior distribution of the life time of the target and the target-observation assignments given the unassigned data points.

Compared to the birth move in \cite{Oh_et_al_2009}, our birth move allows any number of consecutive mis-detections (note the parameter $p_{m}$) and improves the efficiency of the target-observation assignments.  
Also, our birth move proposes the continuous state components of the new born target which are integrated out in \cite{Oh_et_al_2009}.

\paragraph{Extension and reduction moves}
In this move, we choose one of the $K$ existing targets, and extend its track either forwards or backwards in time. 
The idea of forward extension is outlined as follows, and the backward one can be executed in a similar way.  First decide how long we will extend the target based on $p_s$, and decide the detection at each time for the extended part, based on $p_d$ and the number of clutter at that time. To extend from time $t$ to $t+1$,  if the target is detected,  we assign to it an observation chosen from the clutter at time $t+1$ with a probability inversely proportional to its distance to the predicted (prior) mean of the state at $t+1$. (Here, we mean $g^{-1}_{\theta}(y|x)$ by  the `distance'  between $x\in \mathcal{X}$ and $y\in \mathcal{Y}$.) Then we calculate the Gaussian approximation of the state posterior by applying the unscented transformation \cite{wan2000unscented} 
using the chosen observation. 
The forward extension step is repeated forwards in time until we reach the extension length.
Denote $q_{e, \theta}(\hat{z}_e | z_{1:n},\mathbf{y}_{1:n})$ the probability induced here, where $\hat{z}_e$ consists of the new death time and the observation information of the extended part. 
Then, backwards sample the extended part states $\hat{\mathbf{x}}_{e}$ by Gaussian proposals denoted by $q_{e,\theta}( \hat{\mathbf{x}}_{e}|\hat{z}_e, \mathbf{x}_{1:n}, \mathbf{y}_{1:n})$ that is calculated based on the forward filtering density (the Gaussian approximation of the posteriors) used in proposing  $\hat{z}_e$. Finally, $z'_{1:n}$ and $\mathbf{x}'_{1:n}$ can be obtained similarly to the birth move based on the one-to-one mapping in \eqref{eq:TwoNotations}, and the Jacobian term in  \eqref{eq:TransMC}  is $1$.

The reduction move paired with the extension move is implemented as follows. We randomly choose target $k$ among the $K$ existing targets, then choose the reduction type and the reduction time point, either $t \in \{ t_b^k+1, \ldots, t_d^k-1 \}$ to discard  $\{ t, \ldots, t_{d}^{k}-1 \}$ part of the track, or $t \in \{ t_b^k, \ldots, t_d^k-2 \}$ to discard its $\{ t_{b}, \ldots, t \}$ part. Denote $q_{r, \theta}(z_{1:n} | z'_{1:n})$ the probability induced here.
The acceptance ratio of the extension move is
\begin{equation}
r_{3}(z'_{1:n},\mathbf{x}'_{1:n}; z_{1:n},\mathbf{x}_{1:n}) = \frac{p_{\theta}(z_{1:n}', \mathbf{x}'_{1:n}, \mathbf{y}_{1:n})}{p_{\theta}(z_{1:n}, \mathbf{x}_{1:n}, \mathbf{y}_{1:n})} 
 \times \frac{q_{r, \theta}(z_{1:n} | z'_{1:n})}{q_{e, \theta}(\hat{z}_e | z_{1:n},\mathbf{y}_{1:n})q_{e,\theta}( \hat{\mathbf{x}}_{e}|\hat{z}_{e}, \mathbf{x}_{1:n}, \mathbf{y}_{1:n})} \label{eq: AcptRatio_EM}
 \end{equation}
whose reciprocal is the acceptance ratio for the corresponding reduction move.

Compared to the extension/reduction move in \cite{Oh_et_al_2009}, our extension/reduction move is done in both ways instead of merely forward extension. Also the extension move makes use of the hidden states to add in measurements instead of using the last assigned measurement. Again, the continuous state variables are proposed here instead of being marginalised as in \cite{Oh_et_al_2009}.

\subsubsection{Dimension invariant moves} \label{sec: Dimension invariant moves}
These moves leave the dimension of $\mathbf{X}_{1:n}$ invariant and are dedicated to changing the links between the existing target states at successive times (state move) and the assignments between the target states and measurements (measurement move). The target state values are also modified in order to increase the acceptance rate. These two moves are specially designed here, where the state move can be considered  as certain combinations of the split/merge and switch moves in \cite{Oh_et_al_2009}, while the measurement move corresponds to the update move in \cite{Oh_et_al_2009}, but with more choice of modification to the observation assignment.
The diversity of the modification choice is enhanced by introducing the state variables into the sampling space.

\begin{figure}
\begin{subfigure}{\linewidth}
\centering
\includegraphics[width=.7\linewidth, height=.5\linewidth]{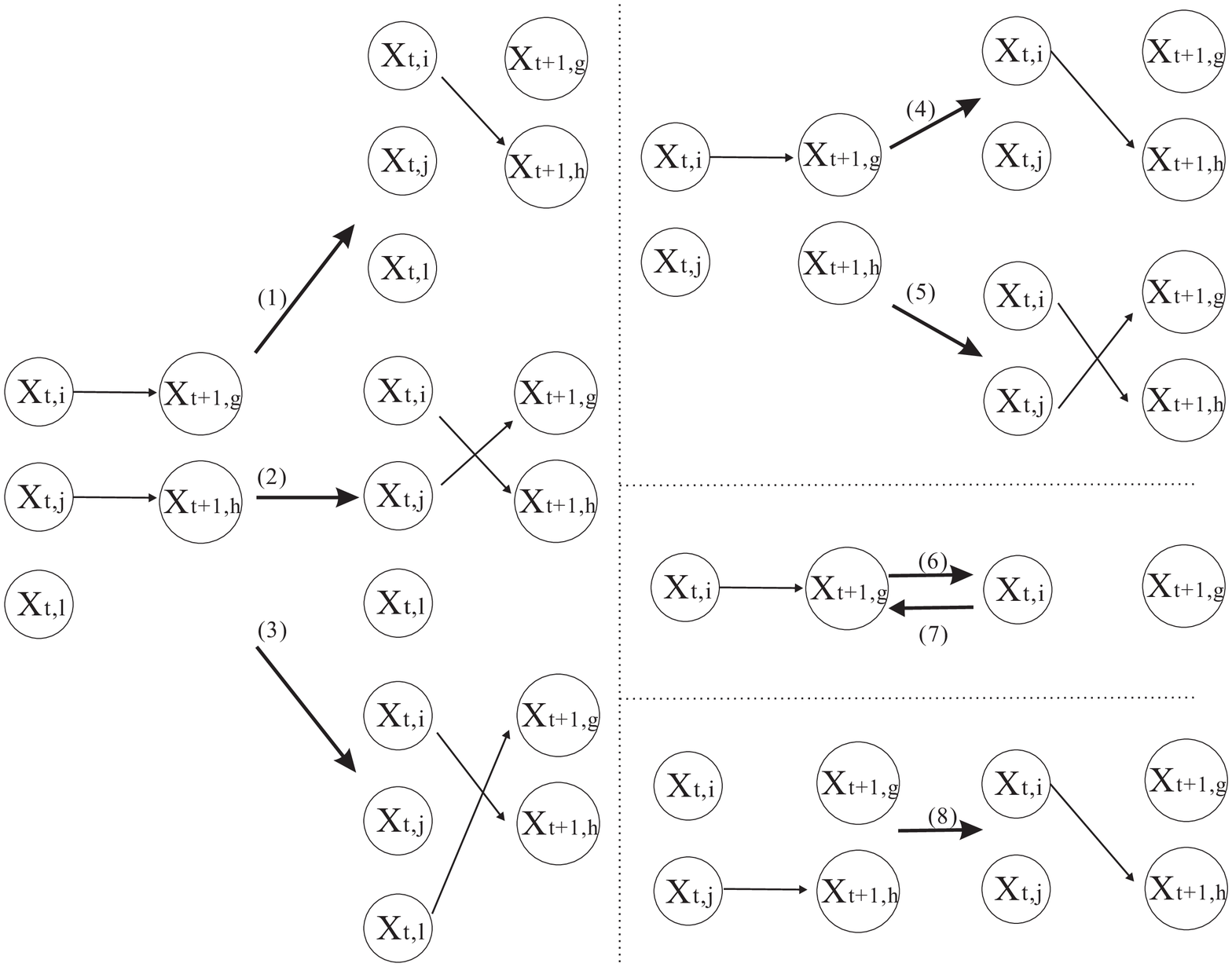}
\caption{state move}
\label{fig:SM}
\end{subfigure}
\begin{subfigure}{\linewidth}
\centering
\includegraphics[width=.7\linewidth, height=.5\linewidth]{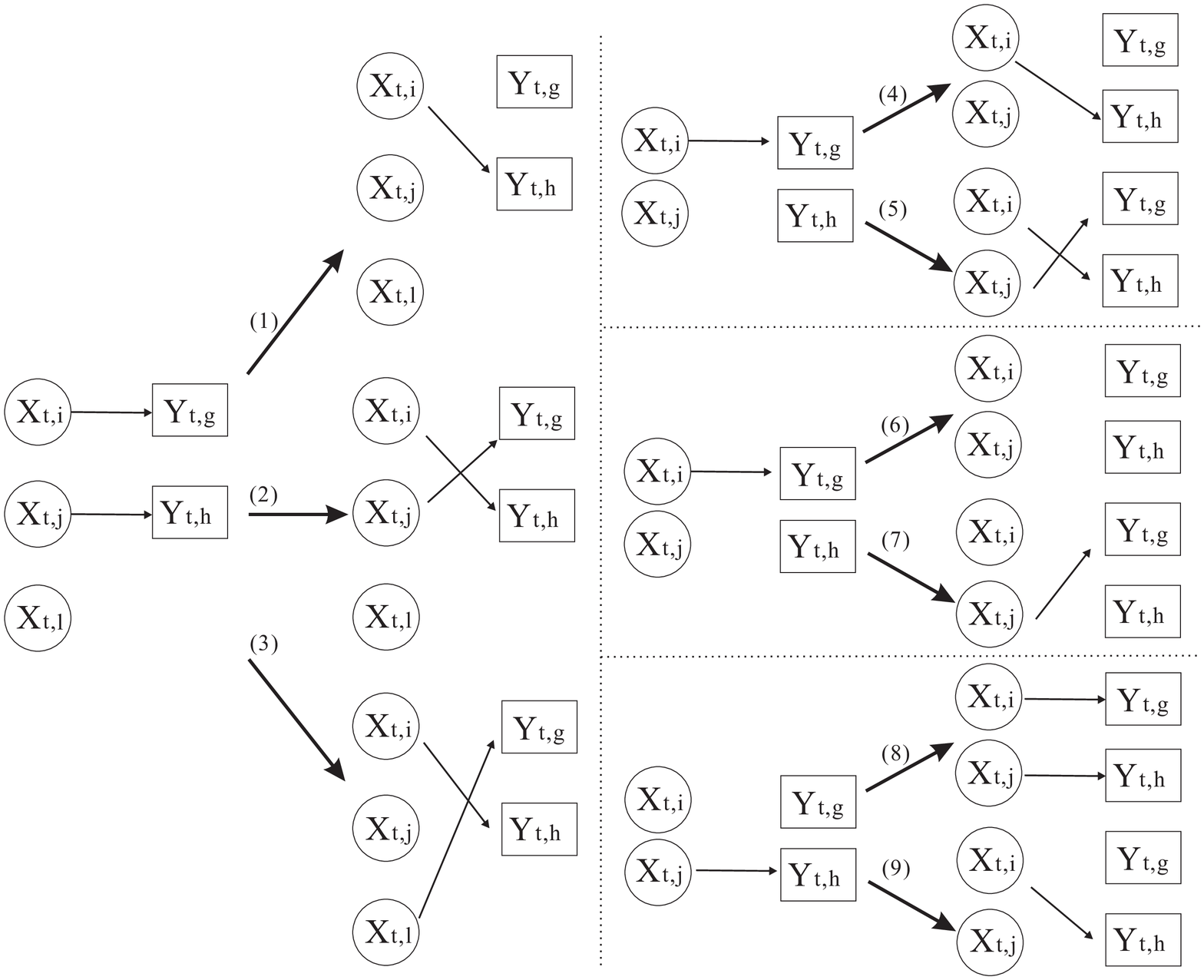}
\caption{measurement move}
\label{fig:OM}
\end{subfigure}
\caption{Graphical illustration of the state move and the measurement move}
\end{figure}

\paragraph{State Move}\label{sec: State move}
In this move,  we randomly choose time $t < n$ and locally change $I_{t+1}^s$, i.e.\ the links between $\mathbf{X}_t$ and $\mathbf{X}_{t+1}$. Figure \ref{fig:SM} is given to illustrate the move. Assume we would like to change the descendant link of $X_{t,i}$. When $X_{t,i}$ has descendant $X_{t+1,g}$,  we can propose to change its descendant to $X_{t+1, h}$ which  originally evolved from $X_{t,j}$ (sub-moves $1, 2, 3$ in Figure \ref{fig:SM}), or to link $X_{t, i}$ to the initial state $X_{t+1,h}$ of a target born at time $t + 1$ (sub-moves $4, 5$), or to delete the link (sub-move $6$). Sub-moves $1, 2, 3$ have different arrangements for the old descendant $X_{t+1,g}$, who becomes clutter in sub-move $1$, or the descendant of $X_{t, j}$ in sub-move $2$ (i.e.\ switches its ancestor with $X_{t+1,h}$), or the new descendant of $X_{t,l}$ in sub-move $3$. Sub-moves $4, 5$ differ in a similar way in terms of the old descendant arrangement. When $X_{t,i}$ has no descendant, it can be merged with a new-born target at time $t+1$ by linking to its initial state (sub-move $7$), or steal another surviving target's descendant (sub-move $8$). Reversibility is ensured by paring sub-moves $1$ and $5$, $6$ and $7$, and the remaining ones with themselves\footnote{For the reversible move of $8$, we choose $X_{t,j}$  to have the descendant link changed. For the other moves, we still choose $X_{t,i}$.}. 
Note that, the new link, e.g, the one between $X_{t,i}$ and $X_{t+1,h}$ in sub-move $1$, means $X_{t+1,h}$ and all its descendants together with their observations will  become $X_{t,i}$'s descendants and the corresponding  observations  in the latter time.  Essentially,  by changing $I_{t+1}^s$ the step described above proposes $\{\hat{z}^{(k)'}, \hat{\mathbf{x}}^{(k)'}\}$ for each target $k$ in set $S$ whose state links are modified.  Denote $q(\hat{z}'_S, \hat{\mathbf{x}}'_S|z_{1:n},\mathbf{x}_{1:n})$ the probability induced here, where $\hat{z}'_S=\{\hat{z}^{(k)'}\}_{k\in S},\,\hat{\mathbf{x}}'_S=\{\hat{\mathbf{x}}^{(k)'}\}_{k\in S}$.

Note that, when the state noise is small, the state move will mostly be rejected if we only modify the state links. Thus, local modification of $\hat{\mathbf{x}}'_S$  is necessary to get state moves accepted. For this reason,  we propose new $\hat{\mathbf{x}}'_{S,w}=\{\hat{\mathbf{x}}^{(k)'}_{w}\}_{k\in S}$, where  $\hat{\mathbf{x}}^{(k)'}_{w}$ is the parts of $\hat{\mathbf{x}}^{(k)'}$ within the time window $w_k = \{ t_{s}^{k}, \ldots, t_{e}^{k} \}$ centred at $t$ with  window size parameter $\tau$ where
\begin{equation}
t_{s}^{k} = \min(t_{b}^{k}, t - \tau + 1), \quad  t_{e}^{k} = \max(t_{d}^{k} - 1, t + \tau),  \label{eq: SMWind}
\end{equation}
by Gaussian proposals, i.e., running UKF and backward sampling for each target $k$ conditioned on its observations in $w_k$ and its states right before and after the window at times $t_{s}^{k} - 1$ and $t_{e}^{k} + 1$ resp., if they exist. 
Denote $q_{s, \theta}(\hat{\mathbf{x}}'_{S,w} | \hat{z}'_{S}, \hat{\mathbf{x}}'_{S}, \mathbf{y}_{1:n})$ the probability density of proposing new local target states.  After updating $\{\hat{z}^{(k)}, \hat{\mathbf{x}}^{(k)}\}$ for each $k\in S$, the unique $z'_{1:n}$ can be obtained by the one-to-one mapping,  and $\mathbf{x}'_{1:n}$ can be obtained by $(\mathbf{x}'_{1:n}, \hat{\mathbf{x}}_{s,w})=\beta_{z_{1:n}, z'_{1:n}} (\mathbf{x}_{1:n}, \hat{\mathbf{x}}'_{s,w})$, which takes out the old states $\hat{\mathbf{x}}_{s,w}$ in the updating windows from $\mathbf{x}_{1:n}$, and inserts $\hat{\mathbf{x}}'_{s,w}$ into $\mathbf{x}_{1:n}$ at the corresponding positions indicated by $z'_{1:n}$. It can be seen that $\beta_{m,m'}$ is invertible with the Jacobian being $1$ as well.  

The acceptance ratio of the state move is
\begin{align}
r_{5}(z'_{1:n},\mathbf{x}'_{1:n}; z_{1:n},\mathbf{x}_{1:n}) = \frac{p_{\theta}(z_{1:n}', \mathbf{x}'_{1:n}, \mathbf{y}_{1:n})}{p_{\theta}(z_{1:n}, \mathbf{x}_{1:n}, \mathbf{y}_{1:n})}\times \frac{q_{s, \theta}(\hat{z}_{S}, \hat{\mathbf{x}}_S | z'_{1:n}, \mathbf{x}'_{1:n}) q_{s, \theta}(\hat{\mathbf{x}}_{S,w} | \hat{z}_{S}, \hat{\mathbf{x}}_{S}, \mathbf{y}_{1:n})}{q_{s, \theta}(\hat{z}'_S,\hat{\mathbf{x}}'_S | z_{1:n},\mathbf{x}_{1:n}) q_{s, \theta}(\hat{\mathbf{x}}'_{S,w} | \hat{z}'_{S}, \hat{\mathbf{x}}'_{S}, \mathbf{y}_{1:n})}\label{eq: AcptRatio_SM}
\end{align}

\paragraph{Measurement Move}
In this move,  we randomly choose time $t$ and locally change $I_{t}^d$, i.e.\ the links between $\mathbf{X}_t$ and $\mathbf{y}_t$. Unlike the state move which modifies $I_{t+1}^s$ followed by modifying local states, the move here first modifies the  states and then proposes the change of $I_t^d$. Specifically, first randomly pick $i\in \{1,\ldots,K_t^x\}$ to decide this move mainly aims at  changing the measurement link of $X_{t,i}$. Assuming the target label of $X_{t,i}$ is $k$,  propose $\hat{\mathbf{x}}'_w$ for target $k$ within the window $w_k =  \{ t_{s}^{k}, \ldots, t_{e}^{k} \} $ 
 similarly  as in the state move, but with the modification to
 disregard the observation of $X_{t,i}$ (if it exists) to remove its influence on $X_{t,i}$. Denote $q_{m, \theta}(\hat{\mathbf{x}}'_w | z_{1:n}, \mathbf{x}_{1:n}, \mathbf{y}_{1:n})$ for the proposal density induced here.
 Then we propose  the change of the measurement link based on the distance between new $X_{t,i}$ and all measurements at time $t$.
Possible proposals are illustrated in Figure \ref{fig:OM} with the similar idea as the state move. First, we set up the measurement link of $X_{t,i}$ if it is not detected (sub moves $8$ and $9$), or choose to modify or delete the measurement link if $X_{t,i}$ is detected (sub moves $1$ to $7$). Then decide how to deal with the original observation if it exists, making it either clutter or new observation of one of the mis-detected targets. Reversibility is ensured by paring sub-moves $1$ and $5$, $6$ and $8$, $7$ and $9$, and the remaining ones with themselves. 
Denote $q_{m, \theta}(z'_{1:n} | \hat{\mathbf{x}}'_w, z_{1:n}, \mathbf{x}_{1:n}, \mathbf{y}_{1:n})$ for the probability induced here.

The acceptance ratio of the measurement move, which is dimension invariant like the state move, can be calculated as
\begin{equation}
r_6(z'_{1:n},\mathbf{x}'_{1:n}; z_{1:n},\mathbf{x}_{1:n})=
\frac{p_{\theta}(z_{1:n}'\mathbf{x}'_{1:n},  \mathbf{y}_{1:n})}{p_{\theta}(z_{1:n} \mathbf{x}_{1:n},\mathbf{y}_{1:n})}\times\\
\hspace{-0.2cm}\frac{ q_{m, \theta}(\hat{\mathbf{x}}_{w} | z'_{1:n}, x'_{1:n}, \mathbf{y}_{1:n}) q_{m, \theta}(z_{1:n} | \hat{\mathbf{x}}_{w}, z'_{1:n}, \mathbf{x}'_{1:n},\mathbf{y}_{1:n})}{ q_{m, \theta}(\hat{\mathbf{x}}'_w | z_{1:n}, \mathbf{x}_{1:n}, \mathbf{y}_{1:n}) q_{m, \theta}(z'_{1:n} | \hat{\mathbf{x}}'_w, z_{1:n}, \mathbf{x}_{1:n}, \mathbf{y}_{1:n})}
\label{eq: AcptRatio_MM}
\end{equation}

\subsection{Update hidden states by particle Gibbs} \label{sec: Update hidden states by particle Gibbs}
Given a joint sample $(z_{1:n}, \mathbf{x}_{1:n})$ obtained via the first loop of  Algorithm \ref{alg: MCMC Tracking}, we may update the target states $\mathbf{x}_{1:n}$ by an MCMC move designed to explore the space of the continuous states. As mentioned in section \ref{sec: Two equivalent notations for MTT}, given $Z_{1:n}$,  $\{\mathbf{X}_{1:n}, \mathbf{Y}_{1:n}\}$ is equivalent to $\{\hat{\mathbf{X}}^{(k)}, \hat{\mathbf{Y}}^{(k)}\}_{k = 1}^{K}$, 
a set of HMMs evolving independently but with the constraint that the target labels need to satisfy the numbering rule.\footnote{More precisely, it is  the numbering of states at each time, which has a one-to-one mapping with the target labels,  that needs to fulfil the numbering rule.} In this move, we do the following: $(1)$ first ignore the labelling constraint, and get new sample $\hat{\mathbf{x}}^{(k)}\sim p_{\theta}(\cdot|\hat{\mathbf{y}}^{(k)})$ independently for each target $k=1:K$; $(2)$ Get a new sample  $(z_{1:n}, \mathbf{x}_{1:n})$  deterministically  from $\{\hat{z}^{(k)}, \hat{\mathbf{x}}^{(k)}\}_{k=1}^K$ by the one-to-one mapping \eqref{eq:TwoNotations} according to the ordering rule.
However, step $(1)$ can not be done directly for non-linear models, so an MCMC move has to be considered. When targets live for long time, prohibitively slow mixing speed prevents us from using MH to update components, even blocks, of $\hat{\mathbf{x}}^{(k)}$. Fortunately, the particle MCMC (PMCMC) framework, in particular particle Gibbs, \cite{andrieu2010particle} provides  an efficient way to update the whole trajectory $\hat{\mathbf{x}}_{k}$ for each $k$ while leaving each $p_{\theta}(\hat{\mathbf{x}}^{(k)} | \hat{\mathbf{y}}^{(k)})$ invariant. The principal idea of PGibbs is to perform a Gibbs sampler on an extended state space whose invariant distribution admits $p_{\theta}(\hat{\mathbf{x}}^{(k)}|\hat{\mathbf{y}}^{(k)})$ as marginal. This can be done by applying a conditional SMC kernel  \cite{andrieu2010particle} for $\hat{\mathbf{x}}^{(k)}$, which is followed by backward sampling \cite{whiteleydiscussion,lindsten2012ancestor}. The application of this idea for the second loop of  Algorithm \ref{alg: MCMC Tracking} is given in Algorithm \ref{alg: MCMC within data association}.

\begin{algorithm}
\DontPrintSemicolon
\KwIn{Current sample $(z_{1:n}, \mathbf{x}_{1:n})$, data $\mathbf{y}_{1:n}$, parameter $\theta$ (in particular, $\psi$)}
\KwOut {Updated sample $(z_{1:n}, \mathbf{x}_{1:n})$}
Extract $\{ \hat{z}^{(k)},  \hat{\mathbf{x}}^{(k)}, \hat{\mathbf{y}}^{(k)}\}_{k = 1}^{K}$ from $(z_{1:n}, \mathbf{x}_{1:n}, \mathbf{y}_{1:n})$ \;
\For{$k = 1:K$}
{Run a conditional particle filter for the HMM $\mu_{\psi}, f_{\psi}, g_{\psi}$ 
with $N$ particles conditional on the path $\hat{\mathbf{x}}^{(k)}$ and the observations $\hat{\mathbf{y}}^{(k)}$; perform backwards sampling 
 to obtain a new sample path $\hat{\mathbf{x}}^{(k)}$.
}
Get updated sample  $(z_{1:n}, \mathbf{x}_{1:n})$ from $\{\hat{z}^{(k)},\hat{\mathbf{x}}^{(k)}\}_{k=1}^K$ by \eqref{eq:TwoNotations} according to  the ordering rule.
\caption{MCMC move to update target states}\label{alg: MCMC within data association}
\end{algorithm}
The PGibbs algorithm with backward sampling has favourable mixing properties, see  \cite{andrieu2010particle, lindsten2012ancestor} for  theoretical analysis and routines of conditional SMC and backward sampling used in Algorithm \ref{alg: MCMC within data association} for a general HMM. We also present the routines in Appendix B.


\section{Static parameter estimation} \label{sec: Static parameter estimation}
In this section, we will show how to extend  Algorithm \ref{alg: MCMC Tracking} to obtain posterior samples of the parameter $\theta$ in the MTT model. To do this, we use the conjugate priors for the components of $\theta$ wherever possible and execute  an MCMC algorithm for $(Z_{1:n}, \mathbf{X}_{1:n}, \theta)$ which is obtained by adding an additional step for sampling $\theta$ to Algorithm \ref{alg: MCMC Tracking} given a joint sample of $(Z_{1:n}, \mathbf{X}_{1:n})$ and the data $\mathbf{y}_{1:n}$. Specifically, starting with an initial $(\theta, z_{1:n}, \mathbf{x}_{1:n})$, we iteratively perform MCMC sweeps given in Algorithm \ref{alg: MCMC for static parameter estimation}.
\begin{algorithm}[h]
\DontPrintSemicolon
\KwIn{Current sample $(\theta, z_{1:n}, \mathbf{x}_{1:n})$, data $\mathbf{y}_{1:n}$, number of inner loops $n_{1}$, $n_{2}$, $n_{3}$}
\KwOut { Updated sample $(\theta, z_{1:n}, \mathbf{x}_{1:n}$)}
\For{$j = 1:n_{1}$}
{Update $z_{1:n}, \mathbf{x}_{1:n}$ by MCMC moves (Algorithm \ref{alg: MCMC across data associations}) to explore $Z_{1:n}$ conditioned on $\theta$.} 
\For{$j = 1:n_{2}$}
{Update $z_{1:n}, \mathbf{x}_{1:n}$ by an MCMC move (Algorithm \ref{alg: MCMC within data association}) to explore $\mathbf{X}_{1:n}$ conditioned on $\theta$} 
\For{$j = 1:n_{3}$}
{Update $\theta$ by an MCMC move conditioned on $z_{1:n}$ and $x_{1:n}$}.
\caption{MCMC for static parameter estimation}\label{alg: MCMC for static parameter estimation}
\end{algorithm}

When we have conjugate priors for all the components of $\theta$, it is possible to implement a  Gibbs move for $\theta$ ($n_{3} = 1$) at the last step of Algorithm \ref{alg: MCMC for static parameter estimation}. Otherwise, one can run an MH algorithm with invariant distribution $p(\theta | z_{1:n},  \mathbf{x}_{1:n}, \mathbf{y}_{1:n})\propto p(\theta)p_{\theta}(z_{1:n}, \mathbf{x}_{1:n}, \mathbf{y}_{1:n})$. In this work, the MTT model used allows us to have a Gibbs move here.

Recall that in \eqref{eq:theta}, $\psi$ is the vector of  HMM parameter for each target, and $p_{s}, p_{d}, \lambda_{b}, \lambda_{f}$ are the parameters governing the data association of the MTT model. Given $z_{1:n}$,  the posterior of  $(p_{s}, p_{d}, \lambda_{b}, \lambda_{f})$ is independent of $\mathbf{x}_{1:n}$ and $\mathbf{y}_{1:n}$, so we refer to them as \textit{data association parameters}. In the following section, we present the conjugate priors and their corresponding posteriors of the data association parameters and the HMM parameters of a non-linear MTT model. 

\subsection{Data association parameters $(p_{s}, p_{d}, \lambda_{b}, \lambda_{f})$} \label{sec: Data association parameters}
Based on the MTT model in section \ref{sec: Multiple target tracking model}, the conjugate priors of $p_{s}, p_{d}, \lambda_{b}, \lambda_{f}$ can be chosen as
\[
p_s, p_d \overset{\text{iid}}{\sim} \text{Unif} (0,1),\quad \lambda_b, \lambda_f \overset{\text{iid}}{\sim}\mathcal{G}(\alpha_0,\beta_0),
\]
where $\text{Unif}(a,b)$ and $\mathcal{G}(\alpha,\beta)$ represent resp. the uniform distribution over $(a,b)$ and the gamma distribution with shape parameter $\alpha$ and scale parameter $\beta$. Note that, we set $\alpha_{0} \ll 1, \beta_{0} \gg 1$ as is commonly done to make the prior less informative,  while a different choice of $\alpha_{0}, \beta_{0}$ can be made when prior knowledge is available. As $K_{t}^{s}$ and $K_{t}^{d}$ are Binomial r.v.'s resp. with success parameters $p_{s}$, $p_{d}$ and number of trials $K_{t-1}^{x}$, $K_{t}^{x}$,  the posteriors distributions of $p_{s}$ and $p_{d}$ are
\begin{align*}
&p_{s} | z_{1:n},\mathbf{y}_{1:n}\sim \mathcal{B} \biggl(1+\sum _{t=1}^n k_t^s,\, 1+\sum_{t=2}^n (k_{t-1}^x-k_t^s)\biggr),\\
 &p_d|z_{1:n},\mathbf{y}_{1:n}\sim\mathcal{B}\biggl(1+\sum _{t=1}^n k_t^d, \, 1+\sum_{t=1}^n (k_t^x-k_t^d)\biggr),
\end{align*}
where $\mathcal{B}(a, b)$ is Beta distribution with parameters $a, b$. As  the number of birth $K_t^b$ and number of clutter $K_t^f$ are Poisson r.v.'s with rates $\lambda_b$ and $\lambda_f$, resp., the posteriors of $\lambda_b, \lambda_f$ are
\begin{align*}
&\lambda_{b} | z_{1:n},\mathbf{y}_{1:n} \sim \mathcal{G} \biggl(\alpha_{0} + \sum_{t=1}^{n} k_{t}^{b},\, (\beta_{0}^{-1} + n)^{-1} \biggr)\\
& \lambda_{f} | z_{1:n},\mathbf{y}_{1:n} \sim \mathcal{G} \biggl( \alpha_{0} + \sum_{t = 1}^{n} k_{t}^{f}, \,(\beta_{0}^{-1} + n)^{-1} \biggr).
\end{align*}

\subsection{HMM parameters $\psi$: an example} \label{sec: HMM parameters: an example}
The choice of conjugate priors of $\psi$ depends on the parametrisation of the HMM model. In the following we adopt the nearly constant velocity model for the state dynamics and the bearing-range model for the measurements as an example. 

\subsubsection{The model} \label{sec: The model}
We assume the state of a target is comprised of its position and velocity in the $xy$ plane, i.e., 
$X=(S_{x}, \dot{S}_{x}, S_{y}, \dot{S}_{y})^T$. The target moves independently in each direction at a nearly constant velocity with the line of sight measurement including the measured range and bearing from  the observer to the target.   The described HMM can be written as follows:
\begin{equation}
X_{t} = F X_{t-1}+ U_{t}, \quad Y_{t} = g(S_{x,t}, S_{y,t}) + V_{t} \label{eq: SigTag}
\end{equation}
with $g: \mathbb{R}^{2} \rightarrow \mathbb{R}^{2}$ defined as
\begin{align}
&g(s_{x}, s_{y}) = \begin{bmatrix} (s_{x}^{2}+ s_{y}^{2})^{1/2} , & \tan^{-1}( s_{y} / s_{x} ) \end{bmatrix}^{T},\nonumber \\
&F = \begin{pmatrix} A & 0_{2 \times 2} \\ 0_{2 \times 2} & A \end{pmatrix}, \quad A =\begin{pmatrix} 1& \Delta \\ 0 & 1 \end{pmatrix} \label{eq: constant velocity}
\end{align}
where $0_{n \times m}$ denotes $n \times m$ matrix of zeros and $\Delta$ is the known sampling interval. The state noise $U_{t}$ and observation noise $V_{t}$ are independent zero-mean Gaussian r.v.'s with covariances $\Sigma_{u}$ and $\Sigma_{v}$ defined as
\begin{equation*}
\Sigma_{u}\!=\!\begin{pmatrix} \sigma_{x}^{2} \Sigma & 0_{2 \times 2} \\ 0_{2 \times 2} & \sigma_{y}^{2} \Sigma \end{pmatrix}\hspace{-0.05cm},\Sigma \!=\! \begin{pmatrix} \Delta^{3}/{3} & \Delta^{2}/{2} \\ \Delta^{2}/{2} & \Delta \end{pmatrix}\hspace{-0.05cm},\Sigma_{v}\!=\!\begin{pmatrix} \sigma_{r}^{2} & 0 \\ 0 & \sigma_{b}^{2} \end{pmatrix}.
\end{equation*}
The initial hidden state is assumed to be Gaussian  distributed with mean $\mu_{b} = (\mu_{bx},0,\mu_{by},0)^T$ and covariance $\Sigma_{b}=\text{diag}(\sigma_{bpx}^2, \sigma_{bvx}^2,\sigma_{bpy}^2, \sigma_{bvy}^2)$. (We set the mean of the initial velocity as $0$ in the absence of more information.)

\subsubsection{Posterior of $\psi$} \label{sec: Posterior of psi}
The parameters of the HMM in the example above are
\[
\psi=(\sigma_x^2, \sigma_y^2, \sigma_r^2,\sigma_b^2, \sigma_{bpx}^2, \sigma_{bpy}^2,\sigma_{bvx}^2,\sigma_{bvy}^2, \mu_{bx}, \mu_{by}).
\]
The priors of all variance  components in $\psi$ are chosen to be inverse gamma distribution with shape parameter $\alpha_{0}$ and scale parameter $\beta_{0}$.
\[
\sigma_x^2, \sigma_y^2, \sigma_r^2,\sigma_b^2, \sigma_{bpx}^2, \sigma_{bpy}^2,\sigma_{bvx}^2,\sigma_{bvy}^2\overset{\text{iid}}{\sim} \mathcal{IG}(\alpha_0, \beta_0).
\]
Again, we can set $\alpha_0 \ll 1,\;\beta_0 \ll 1$ for all to have less informative priors. Given $\sigma_{bpx}^2,\; \sigma_{bpy}^2$, the priors of $\mu_x, \mu_{y}$ are
\[\mu_{bx}|\sigma_{bpx}^2\sim \mathcal{N}(\mu_0,\sigma_{bpx}^2/n_0),\quad \mu_{by}|\sigma_{bpy}^2\sim \mathcal{N}(\mu_0,\sigma_{bpy}^2/n_0)\]
where we can set $n_0$ and $n_0 \mu_0$  small enough to make the prior uninformative. We only discuss the $x$-direction here for the posteriors of the state parameters as the $y$-direction can be deduced in a similar way. For $\sigma_x^2$,  we get the  posterior
\begin{equation*}
\begin{split}
&\sigma_x^2 |\mathbf{x}_{1:n}, z_{1:n}, \mathbf{y}_{1:n}\sim \mathcal{IG}\left(\alpha_0+\sum_{t=1}^n k_t^s,\; \beta_0+\frac{1}{2}\text{tr}(\Sigma^{-1}\hat{\Sigma}^{(x)})\right),\\
&\hat{\Sigma}^{(x)}=\sum_{k=1}^K\sum_{i=1}^{l_k-1} I_x  \left(\hat{x}^{(k)}_{i+1}-F\hat{x}^{(k)}_{i}\right)\left(\hat{x}^{(k)}_{i+1}-F\hat{x}^{(k)}_{i}\right)^{T}  I_x^{T},
\end{split}
\end{equation*}
where $I_x=\begin{pmatrix} 1 & 0 & 0 &0\\0 & 1& 0&0 \end{pmatrix}$.
 For $(\sigma_{bpx}^2, \sigma_{bvx}^2, \mu_{bx})$, denoting $\beta_{1} = \sum_{k=1}^K(\hat{x}^{(k)}_1(1)-\bar{x}_1(1))^2,\; \beta_{2} = \frac{n_{0} K}{n_0+K}(\mu_0- \bar{x}_1(1))^2, \;\beta_{3} = \sum_{k=1}^{K} [ \hat{x}^{(k)}_{1}(2)]^{2}, \quad \bar{x}_1=\frac{1}{K}\sum_{k=1}^K \hat{x}^{(k)}_1$,
 we have
\begin{align*}
&\sigma_{bpx}^2 | \mathbf{x}_{1:n}, z_{1:n}, \mathbf{y}_{1:n}\sim \mathcal{IG}\left(\alpha_{0}+ 0.5 K,\;  \beta_0+ 0.5 (\beta_{1}+ \beta_{2})\right),\\
& \sigma_{bvx}^2|\mathbf{x}_{1:n}, z_{1:n}, \mathbf{y}_{1:n} \sim \mathcal{IG}(\alpha_{0} + 0.5 K,\;\beta_0 + 0.5 \beta_{3}), \\
& \mu_{bx} | \sigma_{bpx}^{2},\mathbf{x}_{1:n}, z_{1:n}, \mathbf{y}_{1:n}\sim \mathcal{N} \biggl( \frac{n_{0} \mu_{0}+K \bar{x}_{1}(1)}{n_{0} + K},\frac{\sigma_{bpx}^{2}}{n_{0} + K} \biggr)
\end{align*}
For measurement parameters $\sigma_r^2$ and $\sigma_b^2$, their posteriors can be obtained by calculating the sum of squared range noise and that of the  bearing noise as follows
\begin{align*}
&\sigma_r^2|\mathbf{x}_{1:n}, z_{1:n}, \mathbf{y}_{1:n}\sim \mathcal{IG}\biggl( \alpha_0+ 0.5 \sum_{t=1}^n k_t^d,\;\beta_{0} + 0.5 \hat{\Sigma}_{v}(1,1) \biggr),\\
&\sigma_b^2|\mathbf{x}_{1:n}, z_{1:n}, \mathbf{y}_{1:n}\sim \mathcal{IG}\biggl(\alpha_0 + 0.5 \sum_{t=1}^n k_t^d,\;\beta_{0}+ 0.5 \hat{\Sigma}_{v}(2,2) \biggr),
\end{align*}
where $\hat{\Sigma}_v=\sum_{t=1}^n\sum_{j: i_{t}^{d}(j) > 0} \Delta y_{t,j}\,(\Delta y_{t,j})^T,\;
\Delta y_{t,j}=y_{t,i_t^d(j)}-g\bigl(x_{t,j}(1), x_{t,j}(3)\bigr)$.

\section{Numerical examples} \label{sec: Numerical examples}
In this section, we give some numerical results to demonstrate the performance of our methods.
All simulations were run in Matlab on a PC with an Intel $i5$ $2.8 \text{ GHZ} \times 2$ processor.

\subsection{Comparison with MCMC-DA for the linear Gaussian model} \label{sec: Comparison with MCMC-DA for the  linear Gaussian  MTT model}
In the linear Gaussian MTT model used here, we assume a target evolves as the first equation in \eqref{eq: SigTag}, but generates observations  according to
\[
Y_{t} = G X_{t} + V_{t}, \; G=\begin{pmatrix}1 & 0&0&0\\0&0&1&0\end{pmatrix},\; V_{t} \sim \mathcal{N}( [0, 0]^{T}, \Sigma_{v}).
\]
The linear observation model is needed by MCMC-DA \cite{Oh_et_al_2009} so that the marginal likelihood $p_{\theta}(\mathbf{y}_{1:n}|z_{1:n})$ can be calculated exactly (i.e, the continuous state variables $\mathbf{X}_{1:n}$ are integrated out). We synthesised data of length $50$ with $p_{s} = 0.95, p_{d} = 0.9, \lambda_{b} = 0.5, \lambda_{f} = 3, \mu_{b} = (80,0,100,0), \Sigma_{b} = \text{diag}(49,9,49,9), \sigma_{x} = 0.7, \sigma_{y} = 1.5, \Sigma_{v} = \text{diag}(4,4)$. 

\begin{figure}
\centering
\includegraphics[width=0.7\linewidth, height=0.4\linewidth]{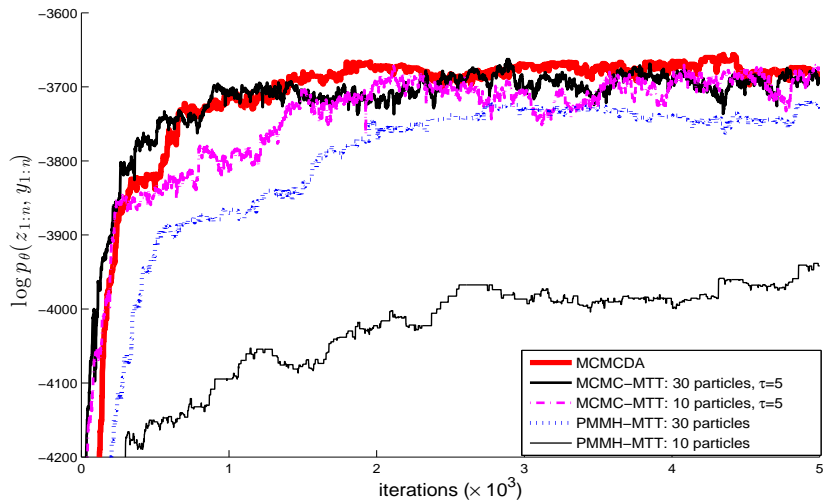}
\caption{Comparisons of Algorithm \ref{alg: MCMC Tracking} (MCMC-MTT) with MCMC-DA \cite{Oh_et_al_2009} and PMMH-MTT \cite{vu2014particle}.  Initial log likelihood is not shown in this zoom in view of  the convergence region}\label{fig:Comparison}
\end{figure}

Assuming all static parameters are known, we compare the performance of  MCMC-MTT in Algorithm \ref{alg: MCMC Tracking},  MCMC-DA  \cite{Oh_et_al_2009} and  PMMH-MTT \footnote{For simplicity, we only implemented the same moves as in the MCMC-DA \cite{Oh_et_al_2009} by substituting the estimate of $p(\mathbf{y}_{1:n}|z_{1:n})$ obtained by particle filters into the acceptation ratio.} \cite{vu2014particle}.  All three methods were initialised by taking all observations as clutter. One iteration of Algorithm \ref{alg: MCMC Tracking} includes a loop of $n_1=50$ MCMC moves to update $(Z_{1:n}, \mathbf{X}_{1:n})$ jointly (with window size $\tau=5$ in moves $5$ and $6$ of Algorithm \ref{alg: MCMC across data associations}), and a loop of $n_2=1$ PGibbs move to update $\mathbf{X}_{1:n}$, while  one iteration of MCMC-DA and PMMH-MTT contains $50$ MCMC moves to update $Z_{1:n}$. Figure \ref{fig:Comparison} shows the plot of $\log p_{\theta} (z^{(i)}_{1:n}, \mathbf{y}_{1:n})$ for the three algorithms, where $z_{1:n}^{(i)}$ is the sample at the $i$-th iteration of each algorithm.
It can be seen that MCMC-DA  outperforms  Algorithm \ref{alg: MCMC Tracking}  with $10$ particles used for each target in the PGibbs step. This is expected since Algorithm \ref{alg: MCMC Tracking}  samples from a larger space $(\mathbf{X}_{1:n}, Z_{1:n})$  than $Z_{1:n}$ alone in MCMC-DA. But its performance almost matches that of MCMC-DA when $30$ particles are used. Two lines at the bottom  of Figure \ref{fig:Comparison} show the performance of PMMH-MTT with $10$ and $30$ particles. We can see that PMMH-MTT converges much slower than Algorithm \ref{alg: MCMC Tracking} especially when the number of particles is small. The slow convergence can be explained by the low acceptance rate (values reported below) due to the high variance of the estimated likelihood.

In terms of computation time, for $10^3$ iterations, MCMC-DA costs $7$ min;  Algorithm \ref{alg: MCMC Tracking} costs around  $7$ (resp. $12$) min for $10$ (resp. $30$) particles per target  (including PGibbs step every $50$ iterations); PMMH-MTT costs around $13$ (resp. $17$) min for $10$ (resp. $30$) particles per target. The average acceptance rate of the MCMC moves that explore the data association  is  about $2.1\%$ for Algorithm \ref{alg: MCMC Tracking} which was  almost the same as MCMC-DA, and $0.48\%$  and $0.3\%$ for PMMH-MTT with $30$ and $10$ particles respectively.

The overall comparison here shows the efficiency of our proposed MCMC moves on the larger sampling space of $(z_{1:n}, \mathbf{x}_{1:n})$:  it can work with much less particles than PMMH-MTT algorithm in \cite{vu2014particle} and can achieve the performance of MCMC-DA \cite{Oh_et_al_2009} within reasonable computation time.

\subsection{Comparison with MHT for the bearing-range model}\label{sec:MHT}
In this experiment, the model described in Section \ref{sec: The model} is assumed and  we set $\sigma_{bpx}^2=\sigma_{bpy}^2=\sigma_{bp}^2, \;\sigma_{bvx}^2=\sigma_{bvy}^2=\sigma_{bv}^2$. Thus,
$
\theta=(p_s, p_d, \lambda_b, \lambda_f, \mu_{bx}, \mu_{by}, \sigma_{bp}^2, \sigma_{bv}^2, \sigma_{x}^2, \sigma_{y}^2, \sigma_{r}^2, \sigma_b^2).
$
We synthesised  data of length $50$ with the parameter vector
 $\theta^*=(0.95, 0.9, 0.4, 3, 80, 100, 64, 9, 0.3, 0.7, 2, 2.5 \times 10^{-3})$
and the sensor located in $[0,0]$ in the window $[-20, 310]\times [-50,210]$ including all the observations inside. The synthetic data used here had  $24$ targets whose trajectories are plotted in the upper half of Figure \ref{fig:NLTrack} where each line of (blue) connected stars  shows connected measurements of one target over the time, and the (red) circles are clutter.
 We compare Algorithm \ref{alg: MCMC Tracking} with the MHT \cite{cox1996efficient} with $L=5$ for $L$-best assignment and $N=3$ for $N$-scan back. To deal with the non-linearity, we replace the Kalman filter in MHT with the unscented Kalman filter \cite{wan2000unscented} which is also used in the MCMC kernels in Algorithm \ref{alg: MCMC Tracking}.
\begin{figure}
\centering
\includegraphics[width=0.8\linewidth,height=0.6\linewidth]{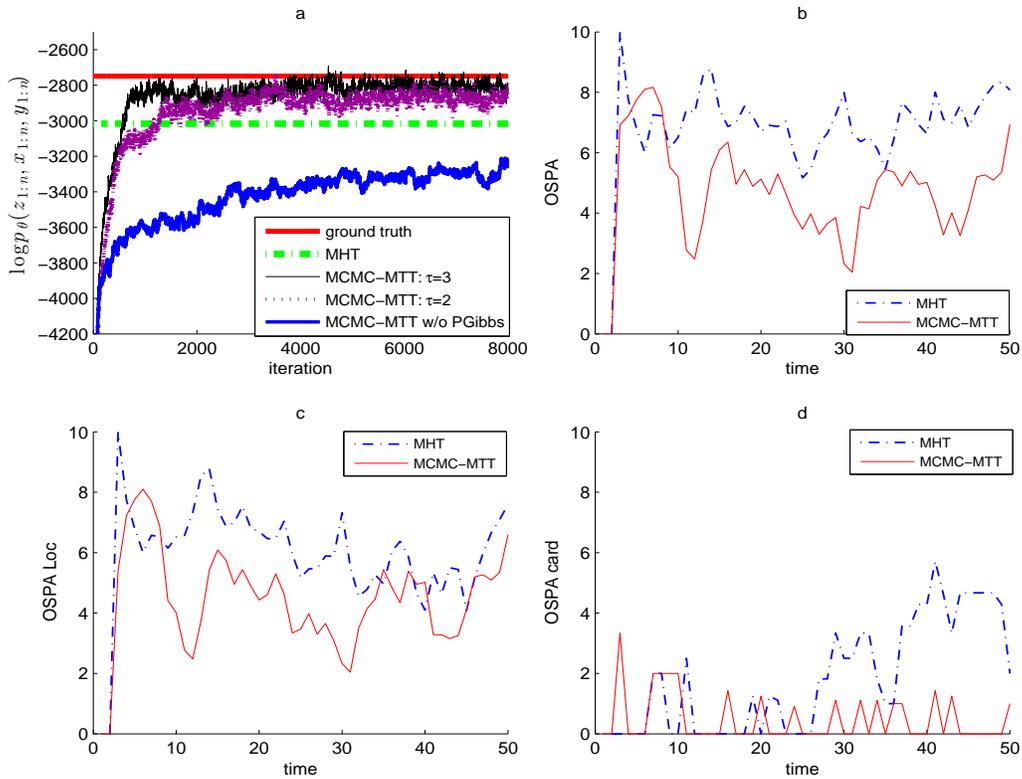}
\caption{Comparison of Algorithm \ref{alg: MCMC Tracking} (MCMC-MTT) and MHT \cite{cox1996efficient}}\label{fig: MHT_Comp_all}
\end{figure}
For the MHT, we ran a particle filter with $300$ particles per target conditioned on the data association output of MHT and perform backwards sampling \cite{doucet2009tutorial} to get  more accurate  state samples. We ran Algorithm \ref{alg: MCMC Tracking} with $15$ particles per target and $n_{1} = 30, n_{2} = 1$. Two window parameters  $\tau=3$ and $\tau=2$ are used  for comparison. In Figure \ref{fig: MHT_Comp_all}(a), we show the joint log-density of $p_\theta(z_{1:n}, \mathbf{x}_{1:n}, \mathbf{y}_{1:n})$ of the output samples of Algorithm \ref{alg: MCMC Tracking}  compared with  the ground truth and the average joint log-density of $p_\theta(z_{1:n}, \mathbf{x}_{1:n}, \mathbf{y}_{1:n})$ obtained using the MHT estimate for $z_{1:n}$ and $500$ SMC samples for $\mathbf{x}_{1:n}$ conditioned on the MHT estimate of $z_{1:n}$. It can be seen that $\log p_{\theta}(z_{1:n}, \mathbf{x}_{1:n}, \mathbf{y}_{1:n})$ for Algorithm \ref{alg: MCMC Tracking} converges to a vicinity of the log-density evaluated at the ground truth, while MHT's output has an apparent gap with the ground truth. We can also see that Algorithm \ref {alg: MCMC Tracking} converges around $1000$ and $1500$ iterations for $\tau=3$ and $2$ resp., which indicates that the mixing speed  of Algorithm \ref {alg: MCMC Tracking} can be improved by the window parameter $\tau$. Additionally, to show the PGibbs step plays a necessary role in Algorithm \ref{alg: MCMC Tracking}, we plot the log-density  obtained by excluding the second loop (PGibbs step) from Algorithm \ref{alg: MCMC Tracking}, which is still far from convergence after $8000$ iterations.

In Figure \ref{fig: MHT_Comp_all}(b), we compare the tracking performance by the OSPA distance \cite{schuhmacher2008consistent}. The OSPA distance is a distance between two sets of points, and it is defined roughly as the sum of a penalty term for the difference in the cardinality of the two sets (OSPA-card) and the minimum sum of distances between the points of those sets (OSPA-loc). These two terms are separately compared in Figure \ref{fig: MHT_Comp_all}(c, d) for Algorithm \ref{alg: MCMC Tracking} and  the MHT algorithm. It can be seen that MCMC outperforms MHT 
in both distances, which agrees with Figure \ref{fig: MHT_Comp_all}(a). However, this better performance comes at the price of longer computation time. For the experiment shown here, $10^3$ iterations of Algorithm \ref{alg: MCMC Tracking} took around $5$ minutes while the MHT took around $1$ minute. Note that, Algorithm \ref{alg: MCMC Tracking} has the potential of  being accelerated by introducing parallel computing techniques which are not used here.

The comparison here shows the better tracking accuracy of Algorithm \ref{alg: MCMC Tracking} over MHT for the non-linear MTT model. The convergence in Figure \ref{fig: MHT_Comp_all}(a)
suggests that Algorithm \ref{alg: MCMC Tracking} is a good choice for batch tracking algorithm for off-line applications. An alternative view would be that the MCMC moves can be used to refine the initial MHT estimate, or that of any other online tracker. Additionally, it also shows the influence of the PGibbs step and the window parameter $\tau$ on the mixing property of the MCMC kernel. The PGibbs step is necessary for the fast mixing property and  we find that setting $\tau$ large than $3$ is normally enough to get good performance.

\subsection{Parameter estimation for the bearing-range model}
\begin{figure}[h]
\centering
\includegraphics[width=0.7\linewidth, height=0.4\linewidth]{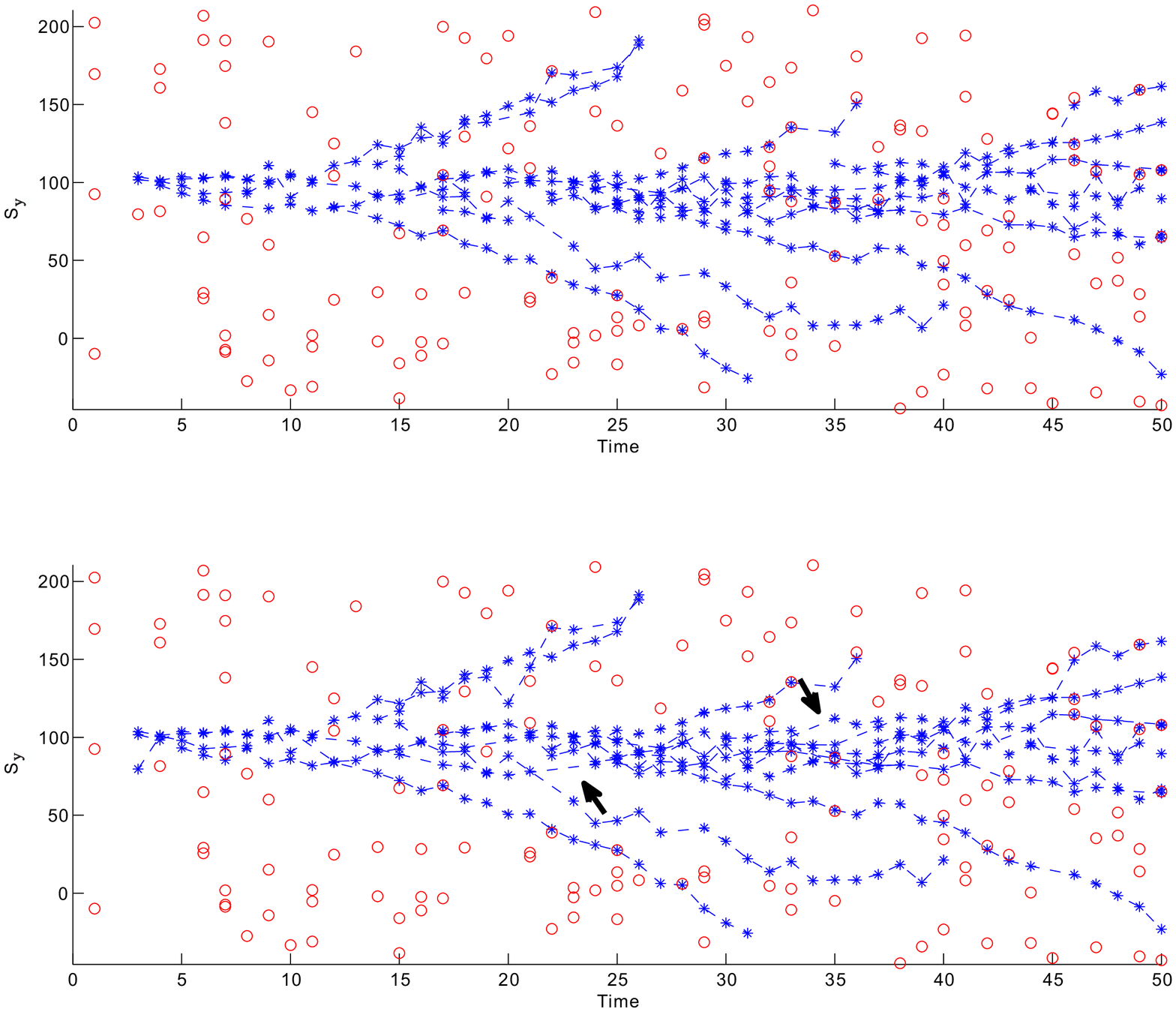}
\caption{Plots of  data association  in the $y$-direction over time: upper figure shows the ground truth and the lower figure posts one sample of data association obtained by Algorithm $4$ in the main paper. Each line of (blue) connected stars shows connected measurements of one target along the time, and the (red) circles are clutter. Arrows (black) indicate where sampled $z_{1:n}$ differs from $z_{1:n}^*$}. 
\label{fig:NLTrack}
\end{figure}

\begin{figure}
\centering
\includegraphics[width=0.8\linewidth,height=0.5\linewidth]{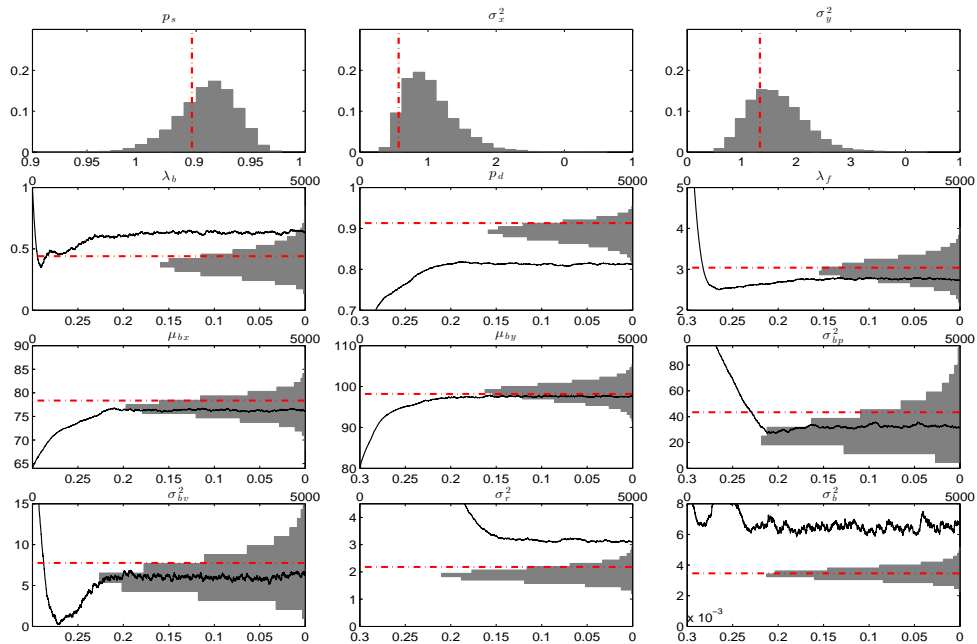}
\caption{Histograms of estimated parameters with (red) dashed line showing the MLE estimate $\theta^{\ast,z,x}$, and the black solid lines showing the estimate using PHD-MLE algorithm in \cite{Singh_et_al_2011} with initial values not shown due to zoom in around $\theta^*$.  Vertical axis shows the parameter value, and bottom horizontal axis shows the normalised histogram height and top horizontal axis shows the iteration step of PHD-MLE algorithm.}\label{fig:NL1}
\end{figure}

Here we demonstrate the joint tracking and parameter estimation performance of  Algorithm \ref{alg: MCMC for static parameter estimation} using  simulated data so that the ground truth is known. The same data set as Section \ref{sec:MHT} is used here.
We initialise $ \theta^{(0)}=(0.6, 0.6, 1, 8, 50, 60, 50, 25, 1, 1.5, 16, 0.02)$, and run $2\times 10^4$ iterations of Algorithm \ref{alg: MCMC for static parameter estimation} with $n_{1}= 60, n_{2} = 1, n_{3} =1$ and $15$ particles. The data association result is shown in  Figure \ref{fig:NLTrack}, the upper half of which is the  ground truth, and the lower half is one sample of  MCMC tracking results.  The histograms of the sampled parameters after $5000$ iterations (burn-in time) are shown in Figure \ref{fig:NL1} where the (red) dashed lines show the MLE estimate $\theta^{*,z,x}$ given the true data association $z^{\ast}_{1:n}$ and true hidden states $\mathbf{x}_{1:n}^*$.  $\theta^{*,z,x}$ is defined as follows.
\begin{equation}
\begin{split}
&(p_s,p_d,\lambda_b,\lambda_f)^{*,z,x} = \arg \max_{p_s,p_d,\lambda_b,\lambda_f} p_{\theta}(z^*_{1:n}), \\
&\psi^{*,z,x}= \arg \max_{\psi} p_{\psi}(\mathbf{x}^*_{1:n},\mathbf{y}_{1:n}|z_{1:n}^*).
\end{split}\label{eq: theta_star}
\end{equation}
Note that, the histograms are an approximation of $p(\theta|\mathbf{y}_{1:n})$. When an uninformative prior is used,  the posterior mode should be consistent with maximum likelihood estimate (MLE) given data $\mathbf{y}_{1:n}$. Since the MLE is not available due to the intractable likelihood, we use $\theta^{*,z,x}$ defined in \eqref{eq: theta_star} instead.

As a final comparison, we compare with the approximate MLE of  $\theta^*$ obtained  by the method in \cite{Singh_et_al_2011}, which proposes to maximise a poisson approximation of $p_{\theta}(\mathbf{y}_{1:n})$ derived similarly as the PHD filter of \cite{Mahler_2003}. We refer to it as the PHD-MLE algorithm. For PHD-MLE, we estimated all the parameters except the survival probability $p_s$, and the state noise  parameters $\sigma_x^2, \sigma_y^2$ (the same as \cite{Singh_et_al_2011}). This is a beneficial setting for the PHD-MLE algorithm as those three parameters are known to it. 
As seen in Figure \ref{fig:NL1}, the PHD-MLE estimates have biases due to the Possion approximation of the data likelihood, especially for the parameters $\lambda_b, p_d, \sigma_r^2, \sigma_b^2$. In computation time, PHD-MLE took $4$ hours to converge (with properly chosen step size), while our method took $40$ min.

\section{Conclusions}\label{sec: Conclusion}
We have proposed a new batch tracking algorithm for the MTT problem with non-linear non-Gaussian dynamics and developed it further for the case when the parameters in the MTT model are unknown. From our experiments, we can see that our MCMC method (Algorithm \ref{alg: MCMC Tracking}) can approach the performance of MCMC-DA \cite{Oh_et_al_2009}, outperforms PMMH-MTT \cite{vu2014particle}, and  obtains  better tracking results compared to MHT \cite{cox1996efficient}. Bayesian estimates of parameters of the non-linear MTT model were also obtained by running  Algorithm \ref{alg: MCMC for static parameter estimation} which includes  an MCMC step for updating the parameters, and outperforms PHD-MLE \cite{Singh_et_al_2011}.

\appendices
\section{Unscented Kalman filter and backwards sampling}\label{Appx: UKF}
Here, we give a short description of the Unscented Kalman filter \cite{wan2000unscented} and backwards sampling \cite{doucet2009tutorial} which are both used in our  MCMC proposals for moves across the data association. The Unscented Transformation (UT) \cite{julier1997new} is a method to calculate the statistics of a random variable undergoing a nonlinear transformations. In UT, a $d$-dimensional random variable $X$ is represented by a set of weighted sigma points $\{\mathcal{X}_i,W_i^m, W_i^c\}_{i=0:2d}$ deterministically chosen to capture its true mean $m_x$ and covariance $P_x$, where
\begin{equation*}
\begin{split}
&\mathcal{X}_0=m_x,\;\mathcal{X}_i=m_x+(\sqrt{cP_x})_i, \; i=1:d,\\
&\mathcal{X}_i=m_x-(\sqrt{cP_x})_i, \; i=d+1:2d
\end{split}
\end{equation*}
$(\sqrt{cP_x})_i$ is the $i$-th row of the matrix square root of $P_x$ multiplied by a scaling parameter $c$  whose value can be set according to \cite{wan2000unscented} together with the mean weight $W_i^m$ and the covariance weight $W_i^c$. After undergoing the nonlinear function $Y=g(X)$, the mean and covariance for Y are approximated by the weighted sample mean and covariance of the transformed sigma points
\[m_y\thickapprox\sum_{i=0}^{2L}W_i^m\mathcal{Y}_i,\quad P_y\thickapprox\sum_{i=0}^{2L}W_i^c(\mathcal{Y}_i-m_y)(\mathcal{Y}_i-m_y)^T\,\]
where $\mathcal{Y}_i=g(\mathcal{X}_i)$. It is proved in \cite{julier1997new} that the estimates of $m_y$ and $P_y$ are accurate to the $3$rd order for Gaussian input and at least the $2$nd order for the other distributed inputs. To introduce UT into filtering,  UKF augments the hidden state $X_t$ to include the state noise and measurement noise, and represents the extended hidden states  by a set of sigma points. The posterior mean and covariance of the hidden state can be obtained by approximating the likelihood  $p(y|x)$  as a Gaussian with  the weighted sample mean and covariance of the transformed sigma points. A detailed description of UKF can be found in \cite{wan2000unscented}.

Using the Gaussian approximations $\{\pi(x_{1:t}|y_{1:t})\}_{t=1:n}$ produced by UKF,   we can do the backwards sampling to get samples from $\pi(\mathbf{x}_{1:n}|\mathbf{y}_{1:n})$ based on decomposition of the joint smooth density
\begin{equation*}
\pi(x_{1:n}|y_{1:n})=\pi(x_n|y_{1:n})\prod _{t=1}^{n-1}\pi(x_t|x_{t+1},y_{1:t})
\end{equation*}
which suggests to first sample $x_n\sim \pi(x_n|\mathbf{y}_{1:n})$, then for $t=n-1:-1:1$, sample $x_t$ according to
\begin{equation*}
\pi(x_t|x_{t+1},y_{1:t})=\frac{f(x_{t+1}|x_{t})\pi(x_t|y_{1:t})}{\pi(x_{t+1}|y_{1:t})}
\end{equation*}

\section{Particle filter and Conditional particle filter}\label{Appx: PF}
Here, we give a short description of the techniques used in the MCMC move that explores $\mathbf{X}_{1:n}$  i.e., Algorithm $3$.
The particle filter approximates the sequence of posterior densities $\{p(x_{1:t}|y_{1:t})\}_{t\geq1}$ by a set of $N\; (N\geq 1)$ weighted random samples called particles
\[\hat{p}_t(\mathrm{d}x_{1:t}|y_{1:t})=\sum_{k=1}^{N} W_t^k\delta_{x_{1:t}^k} (\mathrm{d}x_{1:t}),\quad W_t^{k}\geq 0,\quad \sum_{k=1}^N W_t^{k}=1\]
where $W_t^k$ is called as the importance weight for particle $\mathbf{x}_{1:t}^k$, and $\delta_{x_0}(dx)$ denotes the Dirac delta mass located at $x_0$. These particles are propagated in time using an importance sampling and resampling mechanism.
At time $t=1$, $x_1^{1:N}$ consist of  $N$ independence samples from $q_1(\cdot)$. To propagate from time $t-1$ to $t$, the pair $(A_{t-1}^{1:N}, X_{t}^{1:N})$ is proposed from
\[\rho(x_{t-1}^{1:N},\mathrm{d}a_{t-1}^{1:N})\prod_{n=1}^N q_{t}(x_{t-1}^{a_{t-1}^n},\mathrm{x}_t^n)\]
conditioned on the value of $X_{1:t-1}^{1:N}=x_{1:t-1}^{1:N}$. Here $A_{t-1}^n,n\in1:N$ is the ancestor index of particle $n$ of time $t$, i.e., $x_{1:t}^n=(x_{1:t-1}^{a_{t-1}^n},x_t^n)$, and $A_{t-1}^{1:N}$ are jointly sampled from the resampling distribution $\rho(x_{t-1}^{1:N},\mathrm{d}a_{t-1}^{1:N})$. The multinomial  resampling is one common choice of resampling, where we have $\rho(x_{t}^{1:N},\mathrm{d}a_{t}^{1:N})=\prod_{n=1}^N W_t^{a_t^n}(x_{t}^{1;N})$, and
\begin{equation*}
W_t^{n}(x_{t}^{1:N})=\frac{w_t^{n}(x_{t}^{1:N})}{\sum_{l=1}^N w_t^{n}(x_{t}^{1:N})},\;
w_t^{n}(x_{t}^{1:N})=\frac{g(y_t|x_t^n)f(x_t^n|x_{t-1}^{a_{t-1}^n})}{q_t(x_{t-1}^{a_{t-1}^n},x_t^n)}.
\end{equation*}
Forward particle filtering can be followed by the backwards simulation in Algorithm \ref{alg:BWSim} which makes use of the approximated  marginal filtering density $\hat{p}(x_{t}|y_{1:t})$ to get a path sample from $\hat{p}(x_{1:T}|y_{1;T})$.

Particle Gibbs (PGibbs) sampler \cite{andrieu2010particle} is a valid particle approximation to the Gibbs sampler of $p(\theta, x_{1:n}|y_{1:n})$ with $\theta$  being some parameter variables of HMM models,  where the step of sampling from $p(x_{1:n}|y_{1:n})$ is done by running a \textit{conditional particle filter} (also called conditional SMC) \cite{andrieu2010particle} shown in Algorithm \ref{alg:CAPF}.  Given one path (say the first path) $\mathbf{X}^1_{1:n}=\mathbf{x}^1_{1:n}$ of a particle filter, the conditional particle filter will repopulate the $N-1$ paths conditioned on the first path. It is suggested in \cite{whiteleydiscussion,lindsten2012ancestor} that better mixing property can be achieved by a conditional particle filter followed by a backwards simulator.
\LinesNumbered
\begin{algorithm}
sample $b_T$ according to the multinomial distribution with parameter vector $(N, W_T^{1:N})$\;
\For {$t=T-1:-1:1$}
{ \lFor {$m=1:N$}
  { calculate $W_{t|T}^m=\frac{W_t^m f(x_{t+1}^{b_{t+1}}|x_t^m)}{\sum_l W_t^l f(x_{t+1}^{b_{t+1}}|x_t^l)}$, given $x_{t+1}^{b_{t+1}}$
  }
  sample $b_t$ according to the multinomial distribution with parameter vector $(N, W_{t|T}^{1:N})$\;
}
\caption{Backward simulator}\label{alg:BWSim}
\end{algorithm}
\begin{algorithm}
 set $X_1^1=x^1_1$, sample $X_1^j \sim q_1(\cdot)$, for $j=2:N$, and calculate $W_1^{1:N}$\;
  \For{$t=2:n$}
  { set $X_{t}^1=x^1_t ,\;A_{t-1}^1=1$, sample $A_{t-1}^{2:N}\sim \rho(.|A_{t-1}^1=1)$\;
    for $j=2:N$, sample $X_t^{j}\sim q_{t}(\cdot | X_{t-1}^{A_{t-1}^j})$\;
    calculate $W_t^{1:N}$\;
  }
 \caption{Conditional particle filter}\label{alg:CAPF}
\end{algorithm}





%
\small
\bibliographystyle{IEEEtran}
\bibliography{myrefs_paper}

%
%
%

\end{document}